\documentclass[acmsmall]{acmart}

\usepackage{multirow}
\usepackage{algorithm}
\usepackage{algpseudocode}
\usepackage{bm}

\definecolor{bgcolor}{RGB}{242, 242, 242}
\usepackage{caption}
\usepackage{marvosym}
\newsavebox{\promptboxbox}

\AtBeginDocument{%
  }

\setcopyright{acmlicensed}
\copyrightyear{2024}
\acmYear{2024}
\acmDOI{XXXXXXX.XXXXXXX}

\acmJournal{TOIS}
\begin{document}

\title{Multi-agents based User Values Mining for Recommendation}

\author{Lijian Chen}
\affiliation{%
  \institution{The University of Queensland}
  \city{Brisbane}
  \state{QLD}
  \country{Australia}
}
\email{uqlche22@uq.edu.au}

\author{Wei Yuan}
\affiliation{%
  \institution{The University of Queensland}
  \city{Brisbane}
  \state{QLD}
  \country{Australia}
}
\email{w.yuan@uq.edu.au}

\author{Tong Chen}
\affiliation{%
  \institution{The University of Queensland}
  \city{Brisbane}
  \state{QLD}
  \country{Australia}
}
\email{tong.chen@uq.edu.au}

\author{Xiangyu Zhao}
\affiliation{%
 \institution{City University of Hong Kong}
 \city{Hong Kong}
 \country{China}}
\email{xianzhao@cityu.edu.hk}

\author{Nguyen Quoc Viet Hung}
\affiliation{%
  \institution{Griffith University}
  \city{Gold Coast}
  \state{QLD}
  \country{Australia}
}
\email{henry.nguyen@griffith.edu.au}

\author{Hongzhi Yin}\authornote{Corresponding author.}
\affiliation{%
  \institution{The University of Queensland}
  \city{Brisbane}
  \state{QLD}
  \country{Australia}
}
\email{db.hongzhi@gmail.com}

\renewcommand{\shortauthors}{Chen et al.}

\begin{abstract}
Recommender systems have rapidly evolved and become integral to many online services. However, existing systems sometimes produce unstable and unsatisfactory recommendations that fail to align with users' fundamental and long-term preferences. This is because they primarily focus on extracting shallow and short-term interests from user behavior data, which is inherently dynamic and challenging to model.
Unlike these transient interests, user values are more stable and play a crucial role in shaping user behaviors, such as purchasing items and consuming content. Incorporating user values into recommender systems can help stabilize recommendation performance and ensure results better reflect users' latent preferences. However, acquiring user values is typically difficult and costly.
To address this challenge, we leverage the strong language understanding, zero-shot inference, and generalization capabilities of Large Language Models (LLMs) to extract user values from users' historical interactions. Unfortunately, direct extraction using LLMs presents two key challenges: (1) users often have extensive interaction histories, and the substantial content of each item can easily exceed LLM input length limitations, and (2) LLMs inherently suffer from hallucinations, especially when handling complex tasks.
To overcome these issues, we propose ZOOM, a zero-shot multi-LLM collaborative framework for effective and accurate user value extraction. In ZOOM, we apply text summarization techniques to condense item content while preserving essential meaning. To mitigate hallucinations, ZOOM introduces two specialized agent roles: evaluators and supervisors. The evaluators generate an initial list of user values based on historical interactions, while the supervisors refine these values through a debate process, ensuring consensus and reliability.
After extracting user values, we explore various fusion techniques, such as direct concatenation and contrastive learning, to incorporate these values into recommendation models. Extensive experiments on two widely used recommendation datasets with two state-of-the-art recommendation models demonstrate the effectiveness and generalization of our framework in automatic user value mining and recommendation performance improvement.
\end{abstract}


% \ccsdesc[500]{Security and privacy~Web application security}



\maketitle

\section{Introduction}\label{sec_introduction}
In the era of rapid data growth and information overload, recommender systems have become indispensable for providing personalized suggestions to users across various web applications, such as online news~\cite{ao2021pens}, e-commerce~\cite{schafer1999recommender}, and entertainment~\cite{harper2015movielens}. Over the years, recommender systems have garnered significant attention from both academia and industry, leading to remarkable progress in improving recommendation quality~\cite{yu2023self}. For instance, some researchers have explored a wide range of machine learning techniques, from simple multilayer perceptrons (MLPs)~\cite{he2017neural} to advanced large language models (LLMs)~\cite{yuan2023go,ren2024easyrec}, as the foundation for recommendation models. Others have investigated different ways of structuring user historical data, such as matrices~\cite{koren2009matrix}, graphs~\cite{yu2022graph}, and sequences~\cite{yuan2024fellas}.

Despite these advancements, existing recommender systems often struggle with instability and fail to provide satisfactory recommendations that align with users' fundamental and long-term preferences. A primary reason for this limitation is their reliance on user behavior data to model shallow and short-term interests, which tend to be highly dynamic, sparse, and noisy. Consequently, recommendations generated based on such transient signals may not consistently reflect users' deeper preferences, leading to suboptimal user experiences. For example, a user may accidentally click on items they actually dislike due to distraction or are just interested in at that moment due to curiosity. The recommender system may interpret this as a positive signal and continue suggesting irrelevant content, thereby degrading the user experience.

Unlike shallow and short-term interests, user values represent stable and intrinsic factors that influence decision-making behaviors, such as purchasing items or consuming content. Because of their long-term nature, incorporating user values into recommender systems has the potential to stabilize recommendation performance and generate more meaningful suggestions. For instance, a user might mistakenly click on a news article about a violent crime due to instant interest, even though their core personal value is benevolence. If the recommender system can recognize this underlying value, it can avoid repeatedly suggesting similar content to the user. 

However, obtaining user values is inherently challenging, as they are typically not explicitly available. In practice, user values are often gathered through surveys or questionnaires administered before users engage with a service. However, this approach is costly and time-consuming, requiring well-trained assessors to design questionnaires. Additionally, the engagement rate of users to do the time-consuming surveys is usually low, and self-reported answers may not always reliably reflect users' true preferences. Given the high cost and effort associated with obtaining user values through surveys or external sources, this paper explores an alternative approach: effectively inferring user values from users' historical interaction behaviors. 

Recent advancements in LLMs have demonstrated remarkable capabilities in language understanding, zero-shot inference, and generalization. Recognizing this potential, researchers have increasingly leveraged LLMs to uncover latent patterns from contextual input~\cite{treutlein2024connecting, burns2022discovering}.
However, using LLMs to infer user values from interaction data remains underexplored and presents several challenges. 
Firstly, there is no unified textual framework to describe and for LLMs to generate user values from historical data.
Besides, the length of user interaction histories and the detailed content of each item often exceed LLM input length limitations, making it difficult to process all relevant information effectively.
Moreover, prior studies~\cite{huang2023survey, ji2023survey, maynez2020faithfulness} have shown that LLMs are prone to hallucinations, generating unfaithful or nonsensical outputs, particularly in complex inference tasks like user value extraction.

\begin{figure*}
    \centering
    \includegraphics[width=\linewidth]{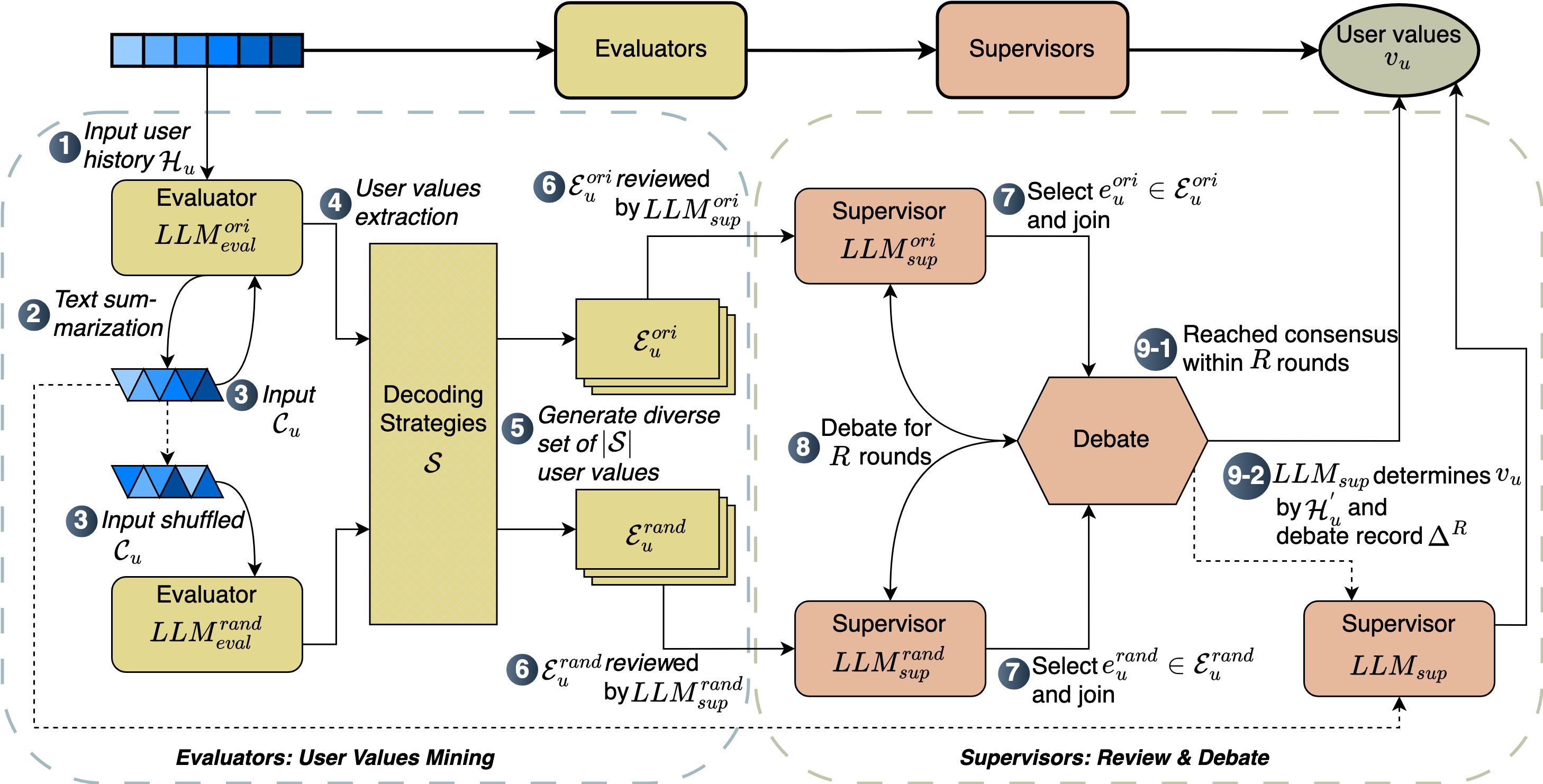}
    \caption{ZOOM Overview}
    \label{fig_zoom_overview}
\end{figure*}

To address these challenges, we propose \textbf{ZOOM} (\underline{Z}ero-sh\underline{o}t Multi-LLMs C\underline{o}llaborative Framework for User Values \underline{M}ining), a novel framework for extracting user values from users' historical interactions, as shown in Figure~\ref{fig_zoom_overview}. To establish a structured framework for value extraction, we adopt Schwartz's Theory of Basic Values~\cite{schwartz2012overview}, which defines ten universal values recognized across cultures~\cite{brandt2017predicting, jaskolka1985measuring}, providing a robust foundation for describing user values. We present the ten universal values from Schwartz's Theory of Basic Values in Section~\ref{sec_user_values}.
To overcome LLM input length limitations, we employ a text summarization approach that condenses item content while preserving its core meaning, enabling the LLM to process users' complete interaction histories efficiently. 
To mitigate hallucinations, ZOOM employs a multi-agent collaboration strategy, incorporating evaluators who generate initial user value predictions and supervisors who refine these predictions through a debate process, ensuring accuracy and consensus.
These evaluators and supervisors are designed with diverse LLM models, such as Llama~\cite{dubey2024llama} and Gemma~\cite{team2024gemma}, and therefore, can generate, evaluate, and discuss with different background knowledge, improving the result consistency.
Furthermore, ZOOM introduces controlled randomness in interaction history selection and diversification of reasoning paths to enhance consistency and minimize hallucinations, following the principle that multiple correct approaches should lead to the same conclusion.

Once user values are extracted, integrating them effectively into recommender systems is a non-trivial challenge. In this paper, we explore various fusion strategies, including direct concatenation and contrastive learning. The direct concatenation approach simply appends the user value vector to the original user preference vector generated by recommendation models. While straightforward, this method may lead to suboptimal performance because (1) user values could dominate the final representation due to dimensional imbalances, and (2) the user value vector and recommendation model's preference vector may not share the same semantic space. To address these limitations, we design a contrastive learning-based auxiliary task that facilitates a smooth and effective integration of user values into recommendation models.

To evaluate the effectiveness of our proposed methods, we conduct extensive experiments on two benchmark datasets: PENS~\cite{ao2021pens} and MovieLens~\cite{harper2015movielens}, utilizing two state-of-the-art language model-based recommendation systems: MoRec~\cite{yuan2023go} and EasyRec~\cite{ren2024easyrec}. The experimental results demonstrate that integrating user values significantly enhances recommendation performance, leading to more stable and personalized suggestions.

To sum up, the main contributions of this paper are as follows:
\begin{itemize}
    \item We take the first step toward automatically mining user values from historical interactions using large language models (LLMs). To address input length limitations and reduce hallucinations, we propose an agent-based framework named ZOOM, which employs two types of agents working collaboratively to accurately uncover user values.
    \item We further explore how to incorporate these user values as features to enhance recommendation performance through various fusion strategies, including concatenation and contrastive learning.
    \item We conduct extensive experiments to validate the effectiveness of our user value mining framework and fusion methods, as evidenced by notable improvements in recommendation performance.
\end{itemize}

The remainder of this paper is organized as follows. Section~\ref{sec_related_work} reviews related literature such as recommender systems and large language models. Section~\ref{sec_preliminaries} introduces the preliminary research background of this work. Following that, Section~\ref{sec_zoom} presents the technical details of the proposed framework, ZOOM. Section~\ref{sec_user_values_for_rec} describes the integration strategy for incorporating user values into the recommender system. Section~\ref{sec_experiments} provides comprehensive experimental details and results. Finally, Section~\ref{sec_conclusion} concludes the paper.

\section{Related Work}\label{sec_related_work}
\subsection{Recommender Systems}
Over the past decades, recommender systems have significantly evolved and now play a crucial role in providing personalized recommendations. Early recommender systems primarily employed collaborative filtering (CF)~\cite{sarwar2001item}, relying on unique identifiers (IDs) to represent users and items and predicting user preferences based on historical user-item interactions. A widely utilized CF technique is matrix factorization (MF)~\cite{koren2009matrix}, which projects user and item IDs into a shared embedding space and estimates user preferences via the inner product of these embeddings. Advancing beyond traditional methods, neural network-based recommender systems such as Neural Collaborative Filtering (NCF)~\cite{he2017neural} and Latent Relational Metric Learning (LRML)~\cite{tay2018latent} emerged. NCF notably replaced inner-product operations with neural network architectures, improving the modeling of complex, nonlinear user-item interactions. LRML further enhanced recommendation performance by employing Euclidean distance metrics within the embedding space. Despite their effectiveness in modeling collaborative signals from user interaction histories, CF methods often face challenges like the cold-start problem caused by data sparsity.

Another significant research direction involves sequential recommender systems, which explicitly model the temporal order of user interactions. Unlike traditional CF methods that treat interactions as unordered, sequential approaches capture dynamic, short-term preferences from user-item interaction sequences. Early sequential models like FPMC~\cite{rendle2010factorizing} and GRU4Rec~\cite{hidasi2015session} used Markov chains and recurrent neural networks (RNNs) for next-item prediction tasks, respectively. More recently, Self-Attentive Sequential Recommendation (SASRec)~\cite{kang2018self} employed the Transformer~\cite{vaswani2017attention} architecture's self-attention mechanism to identify dependencies within user interaction sequences. SASRec excels by focusing on relevant prior interactions, often outperforming earlier RNN-based models in predicting immediate user intent. However, these sequential methods predominantly rely only on IDs, similarly neglecting valuable content information.

To address the limitations inherent in ID-based models, modality-based recommender systems~\cite{he2016vbpr, li2024multi, yuan2023go} have emerged, integrating side information such as textual descriptions, images, or audio data associated with items. Incorporating additional content modalities effectively mitigates cold-start scenarios where new items lack historical interactions. For instance, Visual Bayesian Personalized Ranking (VBPR)~\cite{he2016vbpr} augments traditional MF with visual features extracted from product images, learning visual dimensions of user preferences to enhance recommendations for cold-start items. Yuan et al.~\cite{yuan2023go} propose a recommendation paradigm that replaces traditional ID-based item embeddings with rich representations derived from raw modality features such as text or images. Instead of using a unique item ID to index an embedding table, MoRec leverages pre-trained modality encoders (e.g., RoBERTa~\cite{liu2019roberta} for text or ViT~\cite{dosovitskiy2020image} for images) to directly generate item embeddings from their semantic content. These modality-aware representations are then fed into a recommendation backbone like SASRec~\cite{kang2018self}, which models user preferences based on sequences of interacted items. Representative models such as MoRec~\cite{yuan2023go} fuse content features with interaction data, enabling effective recommendations for new or sparse items where pure ID-based methods typically struggle. 

\subsection{Large Language Models}
Large Language Models (LLMs) represent a major advancement in the field of natural language processing (NLP), primarily based on transformer~\cite{vaswani2017attention} architectures and trained extensively on large-scale textual datasets. Notable examples include GPT-3~\cite{mann2020language}, PaLM~\cite{chowdhery2023palm}, and LLaMA~\cite{touvron2023llama}. By significantly increasing both data volume and model complexity, these models have developed emergent capabilities such as in-context learning~\cite{mann2020language}, chain-of-thought prompting~\cite{wei2022chain}, and instruction following~\cite{peng2023instruction}, greatly enhancing their versatility and range of applications.
LLMs are typically trained through three main stages: pre-training~\cite{zhou2023lima}, supervised fine-tuning (SFT)~\cite{zhang2023instruction, chung2024scaling}, and reinforcement learning from human feedback (RLHF)~\cite{christiano2017deep, stiennon2020learning, ouyang2022training}. During pre-training, models learn from vast unlabeled corpora, enabling them to capture deep semantic relationships and develop general language representations. The subsequent SFT and RLHF stages further align the models with human intentions, improving their ability to follow nuanced instructions and interact more naturally with users.
The applications of LLMs~\cite{zhu2023large, li2023adapting, wu2023bloomberggpt, tang2023does} have expanded rapidly beyond traditional NLP tasks such as text generation, multilingual translation, and sentiment analysis. They now extend into domains that require advanced cognitive abilities, including mathematical reasoning, logical analysis, and strategic planning. LLMs are increasingly deployed in fields such as information systems, education, finance, and healthcare, where they support content understanding, personalized recommendations, conversational interactions, and decision-making processes.

Despite their significant advancements, large language models (LLMs) exhibit several well-documented limitations, most notably token limits and hallucinations. Token limits, or context-length constraints, refer to the maximum number of tokens an LLM can process in a single interaction. Models such as GPT-3~\cite{mann2020language} and Llama-3~\cite{dubey2024llama} have relatively modest context windows of approximately 2,048 and 4,096 tokens, respectively. These constraints hinder their ability to maintain coherence over long dialogues or incorporate extensive historical context, making it difficult to process lengthy documents or large sets of user interaction history without truncation or specialized handling.
Hallucinations present another critical challenge: LLMs may generate outputs that appear plausible but are factually incorrect or unsupported by real-world information. Numerous studies~\cite{huang2023survey, ji2023survey, maynez2020faithfulness} have shown that LLMs frequently produce content that is either unfaithful or nonsensical relative to source material. Hallucinations can be broadly classified into two types: (1) Intrinsic hallucinations, where the generated content conflicts with the source—for example, fabrications or distortions based on existing information; and (2) Extrinsic hallucinations, where the generated content cannot be verified by the source—meaning it is neither supported nor contradicted. Overall, hallucinations are a complex and inherent issue in LLMs, often rooted in the data, training processes, and inference mechanisms underlying these models.

\subsection{Large Language Models for Recommendation}
Given the success of LLMs in understanding and generating text, there is a growing interest in leveraging them to improve recommender systems.
Recent works have begun to bridge large language models (LLMs) with recommendation tasks in various ways. Wang et al.~\cite{wang2024llm} propose LLM4Rec, which integrates an LLM with a graph-based recommender model. The key idea is to represent the user–item interaction graph in natural language, enabling the LLM to process it. They design prompt templates that incorporate graph relational information (e.g., a user’s connections to various items) into textual prompts, allowing the LLM to comprehend the graph structure. This approach combines the LLM’s strong contextual understanding with the relational reasoning capabilities of a graph neural network, enabling the model to leverage edge information typically inaccessible to standard LLMs.
In another study, Wei et al.~\cite{wei2024llmrec} introduce LLMRec, which uses LLMs to augment the data input to traditional recommenders. Instead of modifying the model architecture, LLMRec employs the language model to generate additional knowledge and relationships within the recommendation graph.
Ren et al.~\cite{ren2024easyrec} present EasyRec, which avoids relying on a large generative language model at inference time. Instead, EasyRec uses a lightweight collaborative language model that aligns textual information with user behavior data. It introduces a text-behavior alignment framework, where item descriptions or reviews are encoded into a semantic space and aligned, via contrastive learning, with collaborative signals from user interactions. This method effectively incorporates natural language understanding into a collaborative filtering framework. A key advantage of EasyRec is its effectiveness in text-based zero-shot recommendation, allowing it to recommend items with little or no interaction data by leveraging textual content.
Beyond enhancing recommendation algorithms directly, researchers have also explored LLMs as tools for simulating and understanding user behavior. Zhang et al.~\cite{zhang2024generative} propose Agent4Rec, a novel framework that uses LLM-based agents to mimic user interactions with recommender systems. These generative user agents are equipped with a persona (profile), a memory module, and the ability to perform actions such as clicking or providing feedback. Within a simulated environment, the recommender presents a page of recommendations, and the LLM-agent decides how to respond (e.g., which item to click or whether to skip), step by step. This creates an interactive loop that enables online evaluation without real users. Agent4Rec allows researchers to assess recommendation policies by observing agent responses and to investigate phenomena such as the filter bubble effect by tracking the evolution of agent interests over time.

\section{Preliminaries}\label{sec_preliminaries} 
\subsection{Recommendation Task}
In a recommender system, let $\mathcal{U}$ and $\mathcal{I}$ denote the sets of users and items, respectively. User-item interactions is represented by $\mathcal{X} \in \mathbb{R}^{\vert\mathcal{U}\vert \times \vert\mathcal{I}\vert}$, where $\mathcal{X}^{u,i} = 1$ indicates a positive interaction (e.g., click or purchase), and $\mathcal{X}^{u,i} = 0$ indicates that user $u$ has not interacted with item $i$ or expressed interest in it. A user $u \in \mathcal{U}$ can be represented by a unique ID $u$ or a user profile. Similarly, an item $i \in \mathcal{I}$ can be represented by a unique ID, an item profile, or modality-based features such as textual descriptions. The primary objective of a recommender system is to estimate the probability of a future interaction between user $u$ and item $i$, denoted as $\hat{x}_{u,i}$, and thereby provide personalized recommendations to user $u$. The training objective of the recommender system is defined as:
\begin{equation}
    \underset{\Theta}{\arg\min}\; \mathcal{L}^{rec}({\mathcal{D}} \mid \Theta) \;, \mathrm{where}\; \mathcal{D} = \lbrace (u,i,j) \mid u \in \mathcal{U} \wedge i \in \mathcal{I}_{u}^{+} \wedge j \in \mathcal{I} \setminus \mathcal{I}_{u}^{+} \rbrace
\end{equation}
Here, $\Theta$ denotes the model parameters, $\mathcal{L}^{rec}$ is the recommendation loss (e.g., BPR Loss~\cite{rendle2012bpr}), and $\mathcal{D}$ is the training data.

\subsection{User Values} \label{sec_user_values}  % basic definition & theory of user value
To ensure that recommender systems can reflect the values of the individuals and societies they are designed to serve, we propose integrating user values into the recommendation process. Following~\cite{yao2023value}, we adopt Schwartz's Theory of Basic Values~\cite{schwartz2012overview} as our user values system, given its extensive application in research within economics and political science~\cite{brandt2017predicting, jaskolka1985measuring}. This theory exhibits three crucial characteristics: (1) Clarity, as it provides clear and precise definitions of values, identifying ten universal basic values (organized into four higher-level groups: Openness to Change, Conservation, Self-Enhancement, and Self-Transcendence) and $58$ fine-grained value items; (2) Adaptability, as it can be applied across diverse scenarios and cultural contexts, with its values recognized as universal across cultures; and (3) Transparency, as it enables value-based decision-making processes in AI systems to be more transparent and easier to validate, with values explicitly organized and their relationships (e.g., conflicts between Self-Enhancement and Self-Transcendence) clearly mapped. To be specific, a user's value can include multiple entries from the ten universal basic values, such as Universalism and Benevolence. However, according to the theory, a user value will be considered inappropriate if it includes both Self-Direction and Tradition, due to the conflicts between their corresponding higher-order categories, Openness to Change and Conservation, respectively. This merit can give guidance for judging whether the automatically generated values are suitable or not in this work.

\begin{figure*}
    \centering
    \includegraphics[scale=0.18]{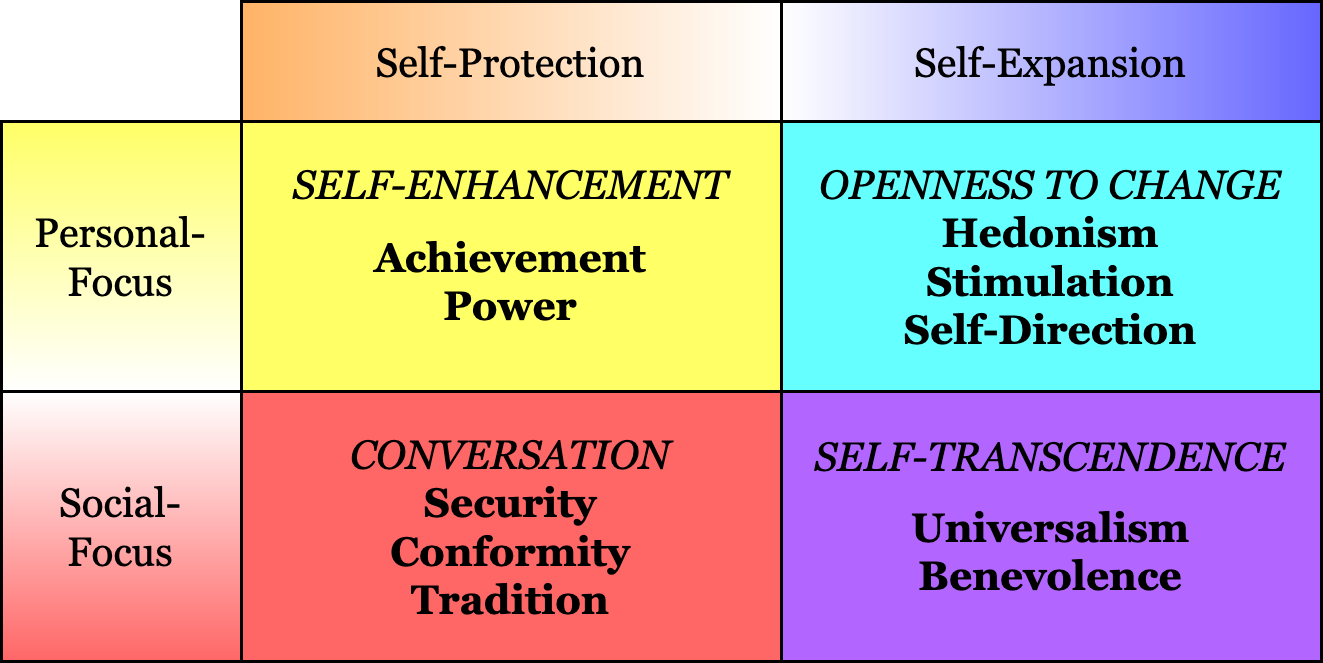}
    \caption{Overview of Schwartz's Theory of Basic Values}
    \label{fig_ten_universal_values}
\end{figure*}

Figure~\ref{fig_ten_universal_values} illustrates the ten universal basic values identified by Schwartz’s Theory of Basic Values (Achievement, Power, Hedonism, Stimulation, Self-Direction, Security, Conformity, Tradition, Universalism, and Benevolence), along with their organization and mapping. In our proposed method, ZOOM, the mined user values are selected from these ten universal basic values. 

\subsection{LLM-based Agents} 
A general LLM-based agent consists of three essential modules: the profile module, the memory module, and the action module.

The profile module defines the fundamental characteristics of an agent, including its role, task, personality traits, domain expertise, and behavioral guidelines. This module customizes the agent's responses and actions to match specific contexts and user expectations.

The memory module is a crucial component that allows the agent to retain and retrieve information from past interactions, supporting coherent and contextually relevant responses. To enable efficient memory management for both storage and retrieval, the memory module is defined as $\mathcal{M} = \left\{ m_{0}, m_{1}, m_{2}, ..., m_{r-1}, m_{r} \right\}$, where $\mathcal{M}$ represents a set of the agent’s responses. In this set, $m_{0}$ typically refers to the profile prompt for the agent, and $m_{r}$ represents the agent’s response at interaction round $r$.

The action module is responsible for implementing the agent’s decisions by converting its internal reasoning into observable actions. During interactions, the agent determines whether to perform a specific action as part of its response, based on a current instruction $\mathcal{P}_{curr}$ and memory $\mathcal{M}$. The action selection process is defined as: $a \sim P(\mathcal{A} \mid \mathcal{P}_{curr}, \mathcal{M})$, where $a$ is the selected action, and $\mathcal{A}$ denotes the set of possible actions.

\section{ZOOM: \underline{Z}ero-sh\underline{o}t Multi-LLMs C\underline{o}llaborative Framework for User Values \underline{M}ining} \label{sec_zoom}
As discussed in Section~\ref{sec_introduction}, user values play a critical role in understanding users' decision-making behaviors in recommendation tasks. However, obtaining user values in practice remains highly challenging. Existing methods typically rely on surveys and questionnaires, which are costly and often suffer from low user participation. Motivated by this limitation, we take the first step toward developing an automatic approach to mine user values. In this section, we introduce the technical details of our carefully designed multi-LLM-based user value mining framework, ZOOM. We begin by outlining the challenges associated with using a vanilla LLM for user value extraction and then describe how ZOOM addresses these challenges through specific technical innovations. The overall architecture of ZOOM is illustrated in Figure~\ref{fig_zoom_overview}, and Algorithm~\ref{algo_zoom} summarizes our method with corresponding pseudocode.

\subsection{Base LLM for User Values Mining}
LLMs have demonstrated remarkable capabilities in language understanding and open-world knowledge reasoning. Naturally, it is tempting to leverage LLMs to extract user values from users' historical interaction behaviors. However, directly applying general-purpose LLMs for user value extraction is non-trivial and presents at least two major challenges.

First, vanilla LLMs are constrained by limited context windows due to architectural and computational restrictions. For instance, Llama-3-8B~\cite{dubey2024llama}, one of the popular LLMs, supports an input length of up to only $4,096$ tokens. Given that interacted items often include extensive metadata, such as titles, attribute descriptions, and main content, such input length limitation severely restricts the LLM's ability to mine values across a broader range of user interaction history.
Second, LLMs are prone to generating unfaithful or counterfactual outputs, a phenomenon commonly referred to as hallucination. This issue becomes even more pronounced when LLMs are tasked with solving complex and advanced problems~\cite{huang2023survey, ji2023survey, maynez2020faithfulness}.
As a result, using a base LLM for user value extraction can only achieve limited performance, as evidenced by the results shown in Table~\ref{tab_ablation}.

\subsection{Multi-agents Collaboration for User Values Mining} \label{sec_user_values_mining}
To address the challenges associated with using LLMs to extract user values from historically interacted items, we propose a user value extraction framework, ZOOM, which is designed as a multi-agent collaboration involving two types of agents: the \textit{evaluator} and the \textit{supervisor}.

The primary role of the evaluator is to generate a diverse set of candidate user values based on a given user's interaction history. To overcome the input length limitation, the evaluator first employs a summarization method to condense the content of interacted items, allowing for more historical interactions to be processed for value generation. To mitigate hallucination, ZOOM utilizes multiple evaluators, each generating user values from different reasoning paths.
Additionally, multiple supervisors are introduced to critically assess the evaluators' outputs and select the most appropriate user values. These selected values are then brought into a debate, where the supervisors collaborate to reach a consensus on the final set of user values.
In the following sections, we provide further details on the roles and processes of these two types of agents.

\subsubsection{Evaluators} \label{sec_evaluators}
Since user value mining depends on users' historically interacted items, which often contain lengthy content, it is inevitable that only a limited number of items can be processed by an LLM due to token constraints. To address this challenge, for a given user $u$, the evaluators are first prompted to summarize the content of the items in the user’s interaction history. They then dynamically truncate the interacted items to ensure that the input remains within the token limit of the LLM. Based on the processed content, the evaluators generate $\vert\mathcal{S}\vert$ candidate values using a set of decoding strategies $\mathcal{S}$.

\begin{figure*}
    \centering
    \includegraphics[scale=0.18]{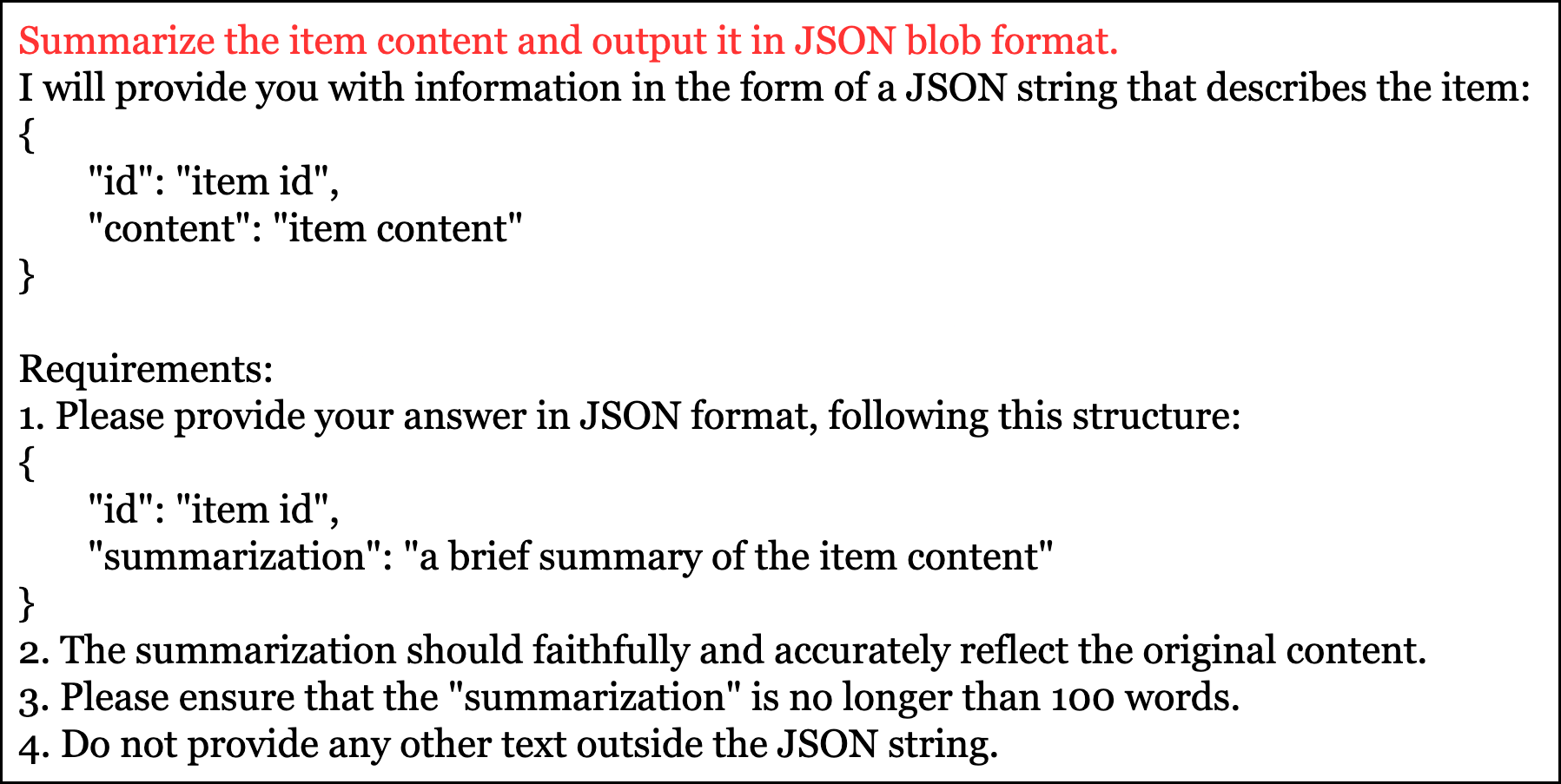}
    \caption{Text Summarization Prompt: $\mathcal{P}_{sum}$}
    \label{fig_p_sum}
\end{figure*}

\textbf{Text Summarization.}
Considering the importance of the number of historically interacted items for user value mining, it is critical to provide as many items as possible to the LLMs to perform user value mining for each user. Inspired by the remarkable performance of LLMs in text summarization~\cite{zhang2024benchmarking}, we prompt the evaluator to summarize the content of the items. This approach compresses the items' content while preserving their original meaning. By doing so, we can include as many historically interacted items as possible within the token limits, enabling more effective user value mining. Formally, the content compression process for an item can be defined as:
\begin{equation}
    c_i = LLM(\mathcal{P}_{sum}, d_i) \;,
\end{equation}
where the $c_i$ denotes the compressed content of the item $i$. $\mathcal{P}_{sum}$ represents the text summarization prompt, which is shown in Figure~\ref{fig_p_sum}. $d_i$ refers to the item's original content.
For the user $u$, after applying summarization to all its interacted items $\mathcal{H}_{u}$, we obtain the compressed set $\mathcal{C}_{u}$. 

\textbf{User Values Extraction.}
Based on the compressed set $\mathcal{C}_{u} = \{c_{i}\}_{i\in \mathcal{H}_{u}}$, the evaluator starts to analyze user values.
Many studies~\cite{yao2023value, zheng2023judging} have suggested that the judgment of LLMs can be easily affected by position bias, causing counterfactual hallucination.
Since user value represents a user's stable and long-term preferences, we do not expect the mined values to be too sensitive to the interacted item orders.
To mitigate this problem, we proactively introduce randomness into the user interaction history input. 
Specifically, we employ two evaluators: one that processes the user's interaction history in its original chronological order, denoted as $LLM_{eval}^{ori}$, and another that works with a randomized order of the user's interaction history, represented by $LLM_{eval}^{rand}$. 
Following the principle that multiple correct approaches should lead to the same conclusion~\cite{wang2022self}, we prompt each evaluator to generate a diverse set of user values for each user through various reasoning paths (i.e., decoding strategies). We use $\mathcal{S}$ to indicate a set of decoding strategies employed to generate a diverse set of user values through various reasoning paths, where $s$ represents one decoding strategy from $\mathcal{S}$. Following Wang et al.~\cite{wang2022self}, we define $\mathcal{S}$ to include four decoding strategies: beam search sampling~\cite{vijayakumar2016diverse}, temperature sampling~\cite{ficler2017controlling}, top-$k$ sampling~\cite{holtzman2018learning, fan2018hierarchical}, and nucleus sampling~\cite{holtzman2019curious}. Consequently, $\mathcal{E}_{u}$ encompasses four user values corresponding to the decoding strategies defined in $\mathcal{S}$.

\begin{figure*}
    \centering
    \includegraphics[scale=0.18]{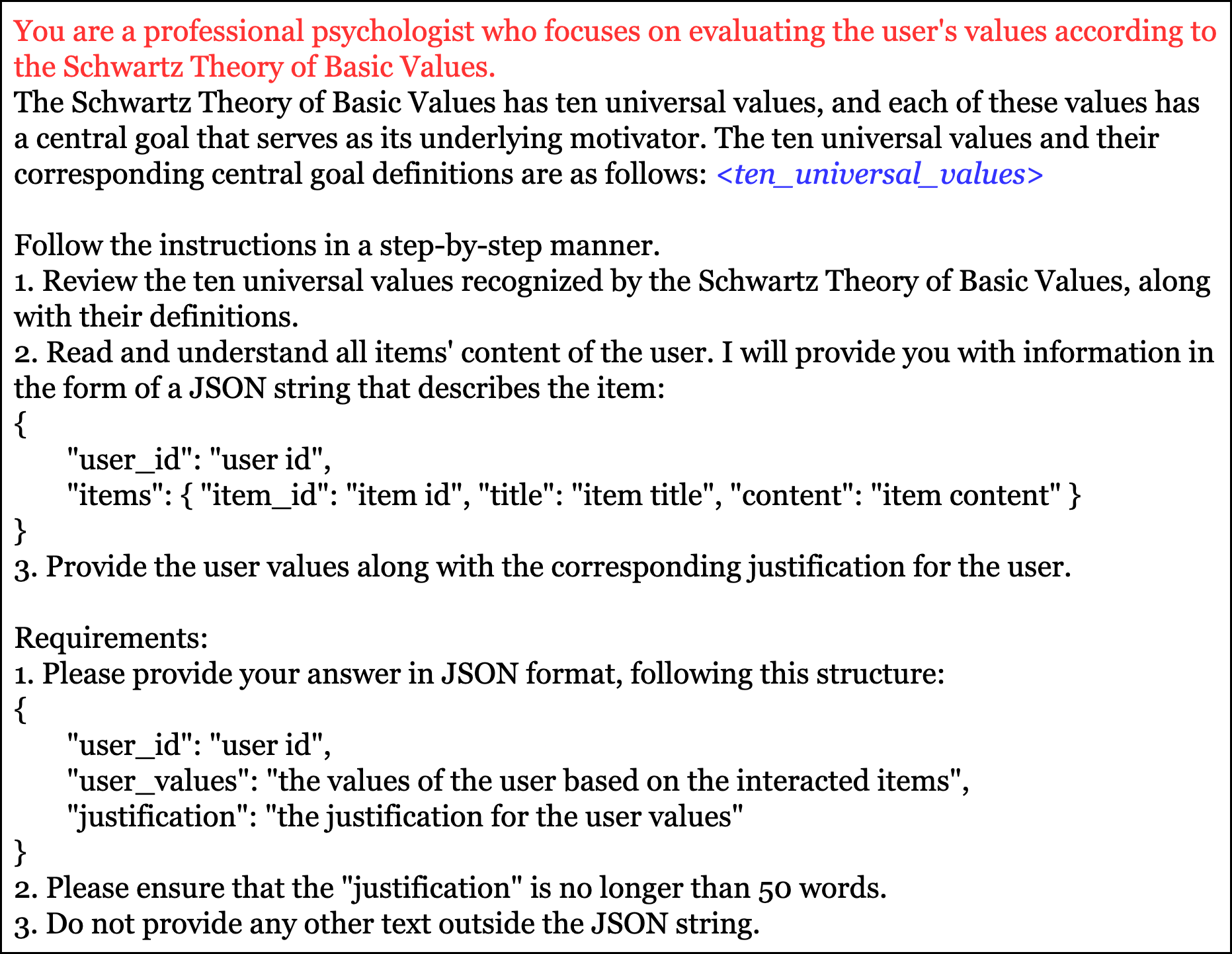}
    \caption{Evaluator Profile Prompt: $\mathcal{P}_{eval}$}
    \label{fig_p_eval}
\end{figure*}

Formally, the process by which an evaluator generates a diverse set of user values for a user $u$ using different decoding strategies can be defined as:
\begin{equation} \label{eq_eval}
    \mathcal{E}_{u}^{k} = \left\{ LLM_{eval}^{k}(\mathcal{P}_{eval}, \mathcal{C}_{u}, s) \mid \forall{s} \in \mathcal{S} \right\} \;, \mathrm{where} \; k\in\lbrace ori, rand \rbrace
\end{equation}
$\mathcal{P}_{eval}$ represents the profile prompt of the evaluator agent, which is presented in Figure~\ref{fig_p_eval}.

\subsubsection{Supervisors} \label{sec_supervisors}
Once both evaluators have completed the user value extraction for user $u$, we introduce a corresponding number of supervisors to review the user values produced. Specifically, each supervisor is paired with one evaluator (denoted as $LLM_{sup}^{ori}$ or $LLM_{sup}^{rand}$) and critically selects the most appropriate user value from the evaluator's generated candidates. Afterward, the two supervisors engage in a debate to reach a consensus on the final set of user values for the given user.

\begin{figure*}
    \centering
    \includegraphics[scale=0.18]{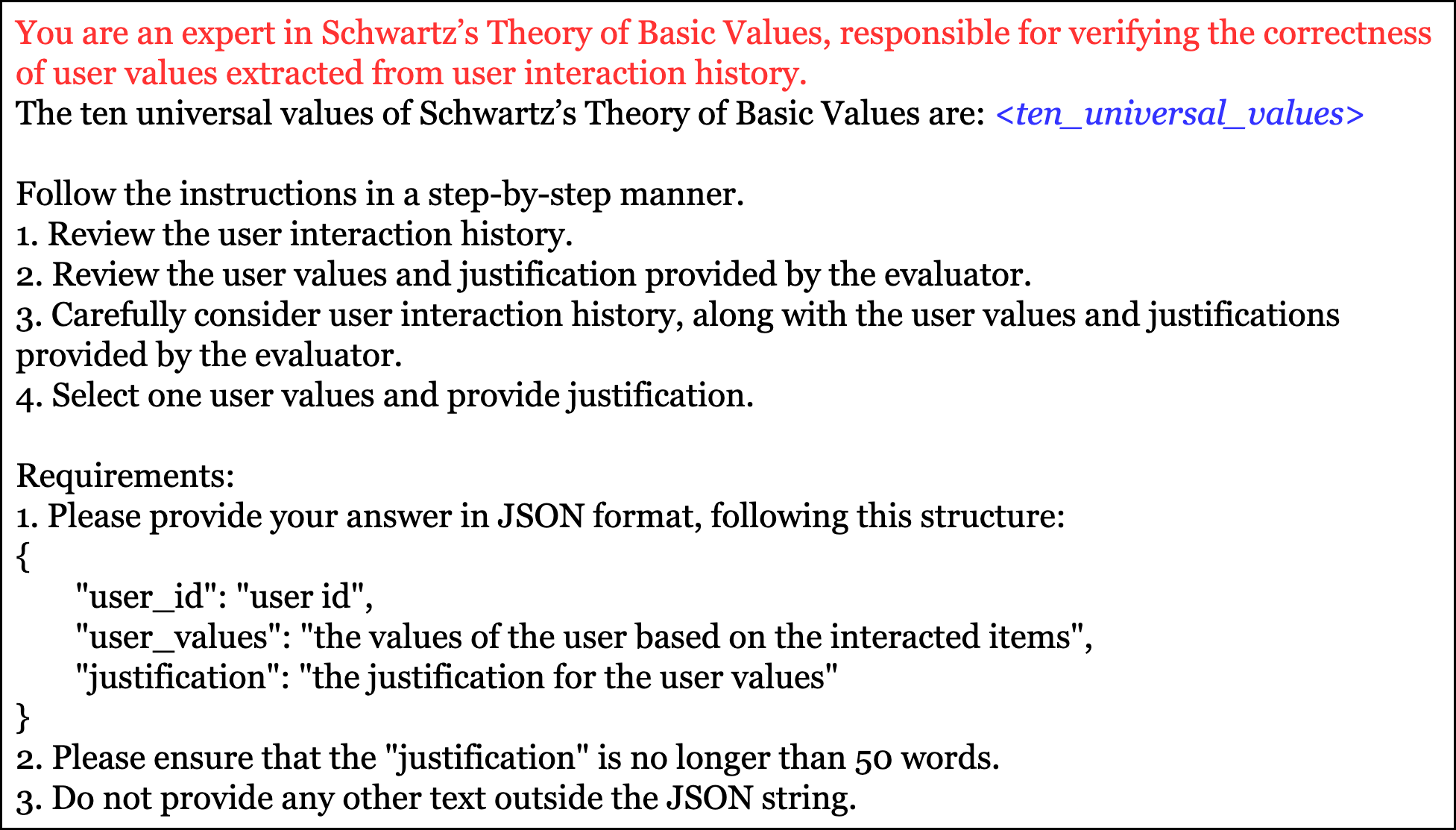}
    \caption{Supervisor Review Prompt: $\mathcal{P}_{review}$}
    \label{fig_p_review}
\end{figure*}

\textbf{Debate.} 
After selecting user values from the evaluators' candidate sets, the two supervisors, $LLM_{sup}^{ori}$ and $LLM_{sup}^{rand}$, engage in a debate process designed to refine and finalize the user values. The debate unfolds over a maximum of $R$ rounds.
Initially, each supervisor independently reviews the candidate values using the review prompt $\mathcal{P}_{review}$ (see Figure~\ref{fig_p_review}), as shown in Equation~\ref{eq_sup_review}.
The outputs of this step are stored in their respective memory states, $\mathcal{M}_{sup}^{ori}[m_{0}]$ and $\mathcal{M}_{sup}^{rand}[m_{0}]$.
\begin{equation} \label{eq_sup_review}
    \mathcal{M}_{sup}^{k}[m_{0}] := LLM_{sup}^{k}(\mathcal{P}_{review}, \mathcal{E}_{u}^{k}) \;, \mathrm{where} \; k\in\lbrace ori, rand \rbrace
\end{equation}

\begin{figure*}
    \centering
    \includegraphics[scale=0.18]{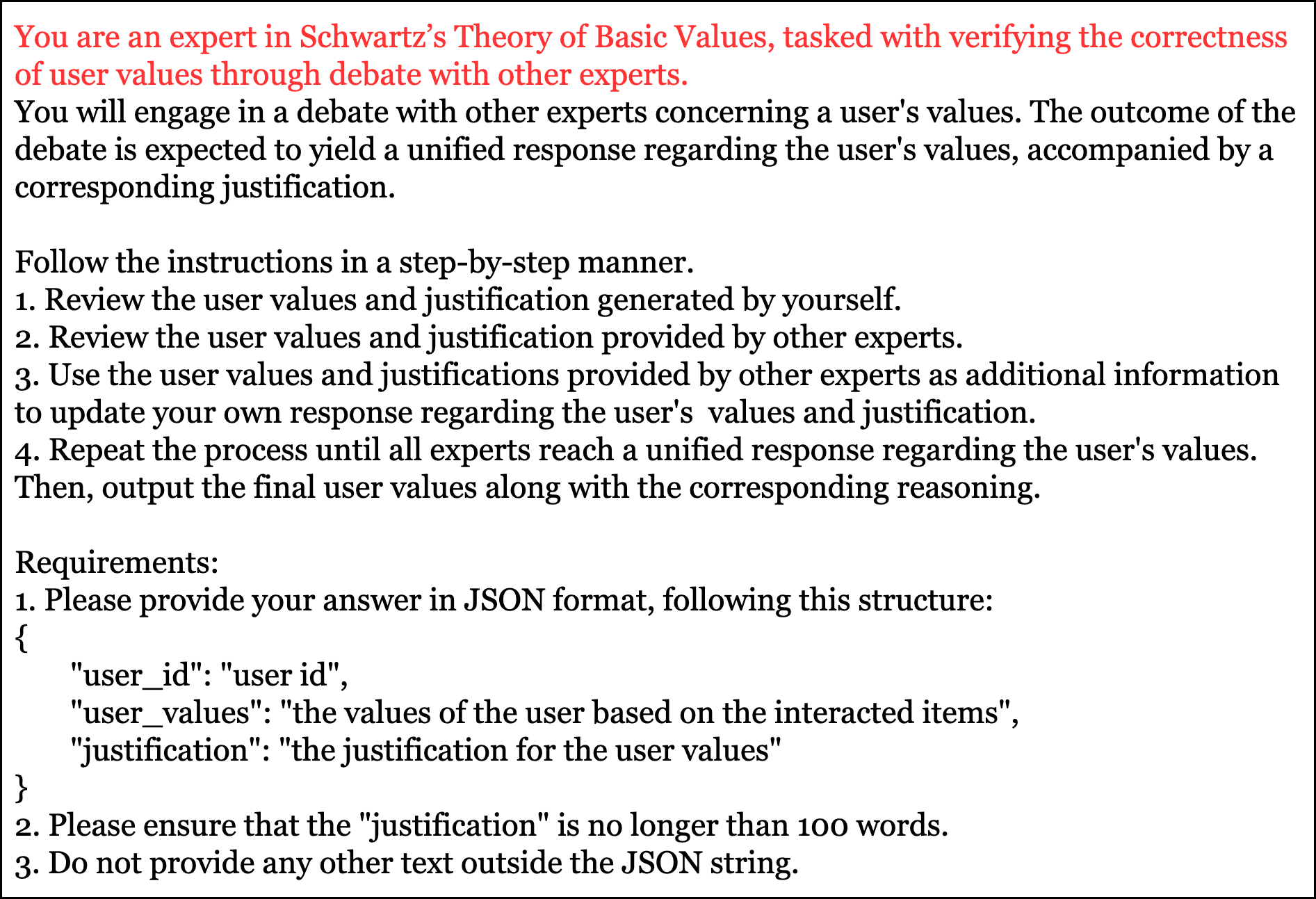}
    \caption{Supervisor Debate Prompt: $\mathcal{P}_{debate}$}
    \label{fig_p_debate}
\end{figure*}

Then, the debate formally begins after this initial review ($r = 0$), where each supervisor submits an initial response based on their memory and evaluation denoted as $\delta_{k}^{0}$.
In each subsequent round $r$ ($1 \leq r \leq R$), supervisors generate updated arguments by considering the debate prompt $\mathcal{P}_{debate}$ (as shown in Figure~\ref{fig_p_debate}), their own model parameters $\psi_{k}$, and the debate history $\Delta^{r-1}$, which consists of both supervisors' previous round responses $\left\{ \delta_{ori}^{r-1}, \delta_{rand}^{r-1} \right\}$.
Therefore, the response generation is governed by the following probability distribution:
\begin{equation} \label{eq_debate_n}
    \begin{aligned}
        P_{sup}(\delta_{k}^{r} \mid \mathcal{P}_{debate}, \psi_{k}, \Delta^{r-1}) \;, \mathrm{where} \; 1 \leq r \leq R; k \in \left\{ ori,rand \right\}
    \end{aligned}
    \end{equation}
As indicated in Equation~\ref{eq_debate_n}, the new response $\delta_{k}^{r}$ depends primarily on the supervisor's internal model parameters $\psi_{k}$, given that both supervisors share the same debate prompt $\mathcal{P}_{debate}$ and history $\Delta^{r-1}$. To promote diversity and prevent collapse to identical outputs, we adopt different backbone LLMs (e.g., Llama~\cite{dubey2024llama}, Gemma~\cite{team2024gemma}) for $LLM_{sup}^{ori}$ and $LLM_{sup}^{rand}$. This architectural variation, along with stochastic decoding strategies, encourages a richer debate dynamic, fostering more thorough reasoning and consensus-building on the final user values.

The debate terminates when one of the following conditions is met: (1) all supervisors reach a consensus, or (2) the debate continues for $R$ rounds without convergence. Du et al.~\cite{du2023improving} argue that, empirically, multiple LLMs are capable of converging to a consensus answer after several rounds of debate. In the case of consensus, a final version of the user values for user $u$ is obtained, i.e., $v_{u}$. However, if the debate reaches the maximum of $R$ rounds without reaching agreement, we introduce a final supervisor as a fallback mechanism. Specifically, this final supervisor reviews the debate history at the final round $R$ and makes the final decision, i.e., it determines the user values for user $u$, as defined in Equation~\ref{eq_debate_final}.
\begin{equation} \label{eq_debate_final}
    v_{u} = LLM_{sup}(\mathcal{P}_{final}, \mathcal{C}_{u}, \Delta^{R}) \;,
\end{equation}
where $v_{u}$ denotes the final user values for user $u$, $\mathcal{P}_{final}$ is the instruction prompt for the final supervisor $LLM_{sup}$, and $\Delta^{R}$ is the debate history at the final round $R$.

\begin{figure*}
    \centering
    \includegraphics[scale=0.18]{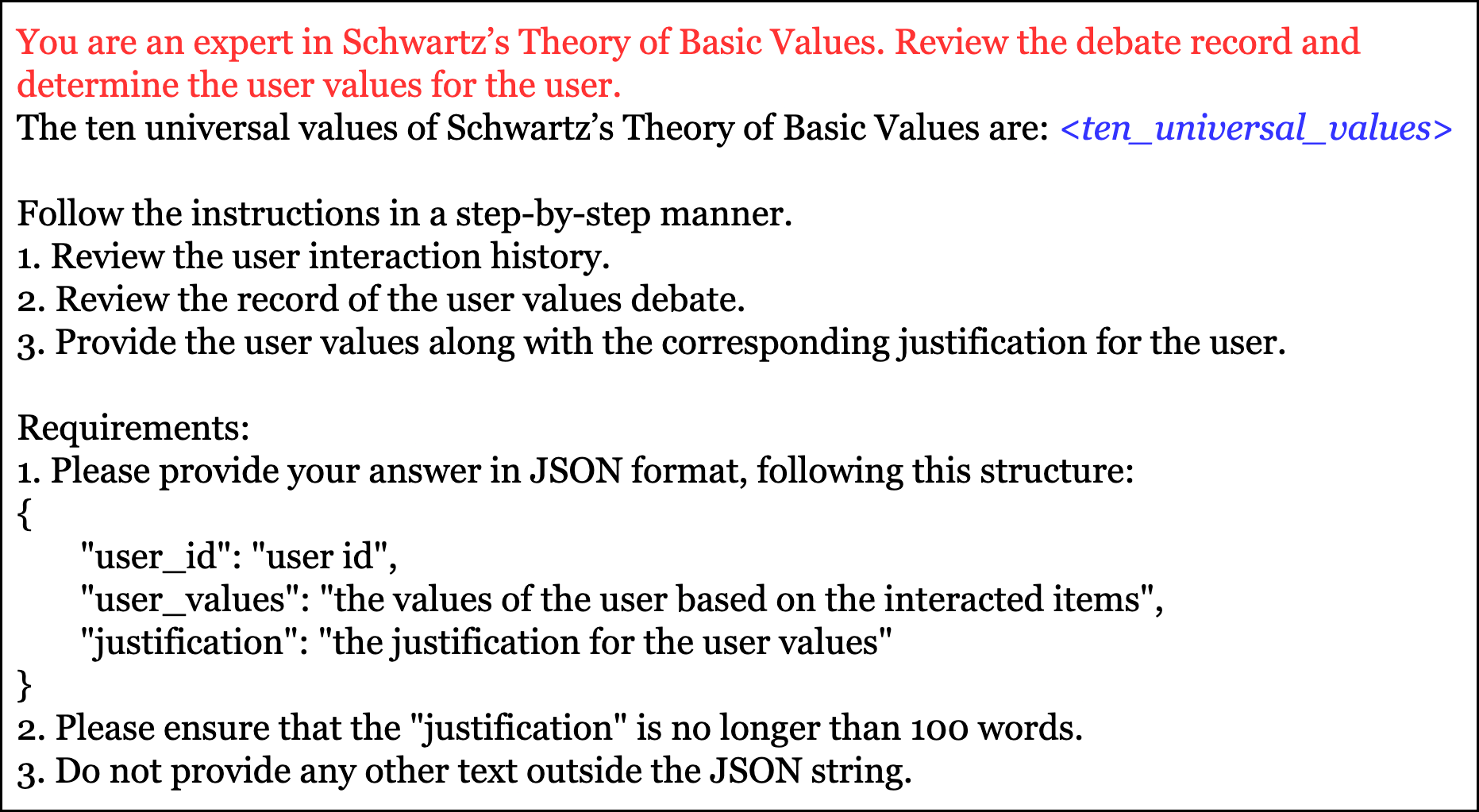}
    \caption{Supervisor Final Determination Prompt: $\mathcal{P}_{final}$}
    \label{fig_p_final}
\end{figure*}

\begin{algorithm}
\renewcommand{\algorithmicrequire}{\textbf{Input:}}
\renewcommand{\algorithmicensure}{\textbf{Output:}}
\caption{ZOOM: \underline{Z}ero-sh\underline{o}t Multi-LLMs C\underline{o}llaborative Framework for User Values \underline{M}ining} \label{algo_zoom}
    \begin{algorithmic}[1]
        \Require 
            \Statex $\mathcal{H}_{u}$: Raw interaction history of user $u$
            \Statex $\mathcal{S}$: Decoding strategies set
            \Statex $\mathcal{P}_{task}$: Instruction prompt, where $task \in \lbrace sum, eval, review, debate, final \rbrace$
            \Statex $LLM_{role}$: LLM agent, where $role \in \lbrace eval, sup \rbrace$
            \Statex $R$: Maximum number of debate rounds
        \Ensure 
            \Statex $v_{u}$: User values of user $u$
        \State $\mathcal{C}_{u} \leftarrow \lbrace LLM(\mathcal{P}_{sum}, d_{i}) \mid d_{i} \in \mathcal{H}_{u} \rbrace $
        \Comment{Summarize the content of all items the user $u$ has interacted with}
        \State $\mathcal{C}_{u} \leftarrow$ Truncate $\mathcal{C}_{u}$ if it exceeds the maximum length else $\mathcal{C}_{u}$
        \For{each $k \in \lbrace ori, rand \rbrace$}
        \Comment{User values extraction by Evaluators}
            \State $\mathcal{C}_{u}^{k} \leftarrow$ Shuffle $\mathcal{C}_{u}$ if $k$ is $rand$ else $\mathcal{C}_{u}$
            \State $\mathcal{E}_{u}^{k} \leftarrow \lbrace LLM_{eval}^{k}(\mathcal{P}_{eval}, \mathcal{C}_{u}^{k}, s) \mid s \in \mathcal{S} \rbrace$
            \Comment{Each evaluator generates user values for user $u$}
            \State $\delta_{k}^{0} \leftarrow LLM_{sup}^{k}(\mathcal{P}_{review}, \mathcal{E}_{u}^{k})$
            \Comment{Each supervisor reviews the user values generated by its corresponding evaluator}
            \EndFor
        \State $\Delta^{0} \leftarrow \lbrace \delta_{ori}^{0}, \delta_{rand}^{0} \rbrace$
        \Comment{Gather the supervisors' reviews to start the debate}
        \For{$r \leftarrow 1$ to $R$}
        \Comment{Start debate for $R$ rounds}
            \For{each $k \in \lbrace ori, rand \rbrace$}
                \State $\delta_{k}^{r} \leftarrow LLM_{sup}^{k}(\mathcal{P}_{debate}, \Delta^{r-1})$
                \Comment{Refer to Eq.~\ref{eq_debate_n}}
            \EndFor
            \State $\Delta^{r} \leftarrow \lbrace \delta_{ori}^{r}, \delta_{rand}^{r} \rbrace$
            \Comment{Document the debate records for round $r$}
            \If{$\delta_{ori}^{r}$ is $\delta_{rand}^{r}$}
            \Comment{Return the user values once all supervisors reach a consensus}
                \State \Return $v_{u} \leftarrow \delta_{ori}^{r}$
            \EndIf
        \EndFor
        \State $v_{u} \leftarrow LLM_{sup}(\mathcal{P}_{final}, \mathcal{C}_{u}, \Delta^{R})$
        \Comment{If no consensus is reached in the debate, a final supervisor determines the user values using $\mathcal{C}_{u}$ and $\Delta^{R}$}
        \State \Return $v_{u}$
    \end{algorithmic}
\end{algorithm}

\section{User Values for Recommendation} \label{sec_user_values_for_rec}
After automatically extracting user values, a key challenge lies in how to effectively integrate these values to improve the performance of recommender systems.
A straightforward approach might be to directly concatenate the user value embeddings with the user preference embeddings generated by the recommendation model. 
However, this naive fusion is often suboptimal and can even harm performance because (1) the user value embedding may dominate the final user representation due to dimensional imbalance, and (2) the semantic spaces of the user preference embedding and the user value embedding may not be naturally aligned.
To address these challenges, we propose a contrastive learning-based~\cite{yu2023xsimgcl} auxiliary task to facilitate a smoother and more effective integration of user values into recommender systems.

Specifically, 
for a given user $u$ and the set of extracted user values $\mathcal{V}$, we treat the user preference embedding $\mathbf{e}_{u_{i}}$ and its corresponding user values embedding $\mathbf{e}_{v_{i}}$ as a positive pair. 
To form negative pairs, we sample $N$ user value embeddings $\left\{ \mathbf{e}_{v_{i1}^{-}}, ..., \mathbf{e}_{v_{iN}^{-}} \right\}$ from the set $\mathcal{V}^{-} := \mathcal{V} \backslash \left\{ v_{i} \right\}$, i.e., user values not associated with user $u$. 
To ensure that the negatives are semantically distinct from the user's actual preferences, we select the $N$ embeddings with the lowest cosine similarity to $\mathbf{e}_{u_{i}}$.
The contrastive learning objective is designed to pull the user preference embedding closer to its corresponding user value embedding, while pushing apart unrelated user value embeddings.
In this way, the user values' semantic information is fused into recommendation's user modeling embedding.
Formally, the contrastive loss is defined as:
\begin{equation} \label{eq_cl_loss}
    \mathcal{L}^{cl} = -\log
    \frac{\exp(\cos(\mathbf{e}_{u_i},\mathbf{e}_{v_i})/\mathcal{T})}
    {\exp(\cos(\mathbf{e}_{u_i},\mathbf{e}_{v_i})/\mathcal{T}) + \sum_{j=1}^{N} \exp(\cos(\mathbf{e}_{u_i},\mathbf{e}_{v_{ij}^{-}})/\mathcal{T})} \;,
\end{equation}
where $\mathcal{T}$ denotes the temperature hyperparameter, $\cos(\cdot,\cdot)$ indicates the cosine similarity. Accordingly, the overall training objective for the recommendation model is defined as shown in Equation~\ref{eq_total_loss}.
\begin{equation} \label{eq_total_loss}
    \mathcal{L} = \mathcal{L}^{rec} + \lambda \mathcal{L}^{cl} \;,
\end{equation}
where $\lambda$ is a hyperparameter that controls the contribution of the contrastive loss $\mathcal{L}^{cl}$, and $\mathcal{L}^{rec}$ denotes the arbitrary recommendation objectives such as BPR Loss~\cite{rendle2012bpr}.

\section{Experiments}\label{sec_experiments}
In this section, we conduct extensive experiments to answer the following research questions (RQs):
\begin{itemize}
    \item \textbf{RQ1:} Can the mined user value improve recommender systems' performance?
    % How effective is the proposed contrastive learning-based auxiliary task in integrating user values into recommender systems?
    \item \textbf{RQ2:} How effective is the proposed automatic user value discovering method (i.e., ZOOM)?
    \item \textbf{RQ3:} How essential is each key component of the proposed method (i.e., ZOOM)?
    \item \textbf{RQ4:} How do hyperparameters affect the integration of user values into the recommender system?
\end{itemize}

\subsection{Experimental Settings}
\subsubsection{Dataset}
To evaluate the effectiveness of our proposed method, we conduct experiments on two widely used real-world recommendation datasets: PENS~\cite{ao2021pens} and MovieLens-1M~\cite{harper2015movielens}. These datasets are well-suited for our task, as they offer rich textual information and a diverse range of item categories.
PENS is a news dataset collected by Microsoft, comprising fifteen categories such as politics, finance, and lifestyle, while MovieLens-1M includes eighteen movie genres, including documentary, drama, and film noir.
For the PENS dataset, following~\cite{zhao2022revisiting, chen2024adversarial}, we filter out users with fewer than ten interactions. After filtering, the dataset contains $8,580$ users, $26,605$ items, and $802,574$ user-item interactions.
For the MovieLens-1M dataset, we remove items without storyline information, resulting in $6,040$ users, $3,670$ items, and $996,861$ user-item interactions.
Following standard practices in implicit feedback recommender systems~\cite{yuan2024ptf}, we binarize user-item interactions by converting all ratings to $r_{ij} = 1$. Negative instances are then sampled at a 1:4 ratio~\cite{he2017neural} for training.
Table~\ref{tab_dataset} summarizes the statistics of the two datasets.

\begin{table}[ht]
    \centering
    \caption{Statistics of PENS and MovieLens-1M}
    \begin{tabular}{ccccc}
        \toprule
        \textbf{Dataset} &\textbf{User\#} &\textbf{Item\#} &\textbf{Interactions\#} &\textbf{Sparsity} \\
        \midrule
        PENS &$8,580$ &$26,605$ &$802,574$ &$99.65\%$ \\
        MovieLens-1M &$6,040$ &$3,670$ &$996,861$ &$95.50\%$ \\
        \bottomrule
    \end{tabular}
    \label{tab_dataset}
\end{table}

\subsubsection{Evaluation Protocol}
We adopt the standard leave-one-out protocol~\cite{magnusson2019bayesian} to construct the training and testing datasets for each user. Specifically, for each user, the most recent interaction is reserved as the test item, while the remaining interactions are used for training. Additionally, the most recent interaction within the training set is selected for validation during each training epoch.
Recommendation performance is evaluated using two widely adopted metrics: Hit Rate at rank K (HR@K) and Normalized Discounted Cumulative Gain at rank K (NDCG@K).
To avoid evaluation bias~\cite{krichene2020sampled}, we compute the metric scores over all items that the user has not interacted with.
Since there is no annotated groundtruth for each user in these datasets, we manually annotate some users and use $F_{1}$ scores to measure the alignment of automatically generated user values with human annotations.

\subsubsection{Baselines}
We select two state-of-the-art recommendation models based on language models, MoRec~\cite{yuan2023go} and EasyRec~\cite{ren2024easyrec}, as our baselines. To demonstrate the effectiveness of our user value mining and integration framework, we compare model performance with and without incorporating user values.
Although both baselines leverage language models, they represent fundamentally different user modeling paradigms. MoRec follows the traditional \textit{ID-based sequential recommendation} approach, whereas EasyRec adopts a \textit{text-based collaborative filtering} strategy. Below, we briefly introduce each model.

MoRec (Modality-based Recommendation)~\cite{yuan2023go} enhances conventional ID-based recommendation by replacing item ID embeddings with rich semantic representations derived from raw modality inputs, such as text and images. It employs pre-trained encoders (e.g., RoBERTa~\cite{liu2019roberta} for textual data and ViT~\cite{dosovitskiy2020image} for visual data) to generate item embeddings, which are then input into a sequential recommendation backbone (e.g., SASRec~\cite{kang2018self}) to capture user behavior over time.

EasyRec~\cite{ren2024easyrec} is a language model-based recommendation framework designed to unify semantic content understanding and collaborative filtering. It excels in zero-shot and cold-start scenarios by transforming both user and item data into rich textual profiles that capture content features and collaborative signals. Item profiles are generated by combining structured metadata (e.g., titles and descriptions) with unstructured user-generated content (e.g., reviews), and then processed by a large language model (LLM) to produce semantically enriched descriptions. User profiles are constructed by aggregating representations of previously interacted items along with corresponding user feedback, which are also synthesized via an LLM to form comprehensive preference summaries. Finally, both user and item profiles are encoded using a multi-layer bidirectional Transformer, and the resulting embeddings are used for recommendation.

\subsection{Implementation Details}
In this section, we present the implementation details of our experiments, including the implementation of base models, the implementation of the automatic user value mining framework (ZOOM), and the implementation of user value integration. 

\subsubsection{Implementation of Base Models}
All base models, namely MoRec and EasyRec, are implemented using PyTorch~\cite{paszke2019pytorch}, and we follow the same settings as described in their original papers~\cite{yuan2023go, ren2024easyrec}. Specifically, for MoRec, we use BERT-base~\footnote{\url{https://huggingface.co/google-bert/bert-base-uncased}} as the modality encoder to process the metadata (i.e., item title) of all items, since MoRec replaces item ID embeddings with modality feature embeddings. The embedding size is set to $512$, the learning rate to $1e-4$, and the weight decay to $0.01$. For EasyRec, we adopt RoBERTa-large~\footnote{\url{https://huggingface.co/FacebookAI/roberta-large}} as the encoder backbone to process both user and item profiles. The learning rate is set to $5e-5$, and the number of diverse profiles is set to $2$.

\subsubsection{Implementation of ZOOM}
ZOOM consists of two primary LLM agents: the evaluator and the supervisor. For the decoding strategies $\mathcal{S}$ used by the evaluator agent, we define four distinct strategies: beam search sampling with the number of beams set to $2$, temperature sampling with temperature set to $0.3$, top-$k$ sampling with $k$ set to $20$, and nucleus sampling (also known as top-$p$ sampling) with $p$ set to $0.5$. Accordingly, each evaluator generates a set of four diverse outputs, with each output corresponding to a different decoding strategy.

For the debate process managed by the supervisor agents, we set the maximum number of debate rounds $R$ to $3$. The number of participants in each debate is two: one representing the response generated by $LLM_{eval}^{ori}$, and the other representing the response from $LLM_{eval}^{rand}$. As discussed in Section~\ref{sec_user_values_mining}, we use different LLMs to promote diversity in the responses. To achieve this, the backbone architecture of the supervisor agent is randomly selected from either Llama-3-8B~\footnote{\url{https://huggingface.co/meta-llama/Meta-Llama-3-8B-Instruct}} or Gemma-2-9B~\footnote{\url{https://huggingface.co/google/gemma-2-9b-it}}, both of which are open-source. In particular, when $LLM_{sup}^{ori}$ is assigned Llama-3-8B, Gemma-2-9B is used for $LLM_{sup}^{rand}$, and vice versa. For efficient inference, we use the vLLM library~\footnote{\url{https://docs.vllm.ai/en/stable/}}~\cite{kwon2023efficient}.

\subsubsection{Implementation of Integrating User Values Into Recommender Systems} We investigate two fusion strategies for integrating user values into recommender systems: direct concatenation and a contrastive learning-based auxiliary task. In the direct concatenation approach, the user value embedding is simply added to the user preference embedding within the recommender system. While this method is straightforward, it often leads to suboptimal recommendation performance due to the mismatch between the semantic spaces of user values and user preferences. To address this issue, we propose a contrastive learning-based auxiliary task that facilitates the smooth and effective integration of user values into recommender systems.
 Specifically, we set the contrastive learning temperature $\mathcal{T}$ to $0.1$, and tune the number of negative samples $N$ from the set $\left\{1, 3, 5, 7, 9\right\}$, as well as the weight $\lambda$ from the set $\left\{0.0001, 0.001, 0.01\right\}$.

 \begin{table}[]
    \centering
    \small
    \setlength\tabcolsep{6.0pt}
    \caption{Recommendation Performance Comparisons on Integrating User Values into Recommender Systems. ``Original'' denotes the use of original data without user values integration; ``DC'' indicates integration of user values via direct concatenation; ``CL'' incorporates user values through a contrastive learning-based auxiliary task. For each dataset, the best-performing result within each base recommender system is shown in bold, and the second-best is underlined. ``Improv.'' represents the improvement of ``CL'' over ``Original''.}
    \begin{tabular}{cccccccc}
        \toprule
        \multirow{2}{*}{Dataset} &\multirow{2}{*}{\begin{tabular}[c]{@{}c@{}}Base\\Recommender System\end{tabular}} &\multirow{2}{*}{Metric} &\multirow{2}{*}{$K$} &\multirow{2}{*}{Original} &\multirow{2}{*}{DC} &\multirow{2}{*}{CL} &\multirow{2}{*}{Improv.\;$\uparrow$} \\
        & & & & & & & \\
        \midrule
        \multirow{8}{*}{PENS}
            &\multirow{4}{*}{MoRec}
                &\multirow{2}{*}{HR@$K$} &10 &0.1226 &\underline{0.1240} &\textbf{0.1280} &4.40\% \\
                & & &20 &0.2005 &\underline{0.2013} &\textbf{0.2037} &1.60\% \\
                \cline{3-8}
                & &\multirow{2}{*}{NDCG@$K$} &10 &0.0161 &\underline{0.0165} &\textbf{0.0170} &5.59\% \\
                & & &20 &0.0204 &\underline{0.0207} &\textbf{0.0211} &3.43\% \\
            \cline{2-8}
            &\multirow{4}{*}{EasyRec}
                &\multirow{2}{*}{HR@$K$} &10 &\underline{0.0824} &0.0816 &\textbf{0.0867} &5.22\% \\
                & & &20 &0.1397 &\underline{0.1400} &\textbf{0.1478} &5.80\% \\
                \cline{3-8}
                & &\multirow{2}{*}{NDCG@$K$} &10 &0.0082 &\underline{0.0084} &\textbf{0.0087} &6.10\% \\
                & & &20 &0.0091 &\underline{0.0093} &\textbf{0.0097} &6.59\% \\
        \midrule
        \multirow{8}{*}{MovieLens-1M}
            & \multirow{4}{*}{MoRec}
                &\multirow{2}{*}{HR@$K$} &10 &0.2586 &\underline{0.2633} &\textbf{0.2725} &5.38\% \\
                & & &20 &\underline{0.3792} &0.3789 &\textbf{0.3812} &0.53\% \\
                \cline{3-8}
                & &\multirow{2}{*}{NDCG@$K$} &10 &0.0393 &\underline{0.0405} &\textbf{0.0415} &5.60\% \\
                & & &20 &\underline{0.0455} &0.0453 &\textbf{0.0460} &1.10\% \\
            \cline{2-8}
            & \multirow{4}{*}{EasyRec}
                &\multirow{2}{*}{HR@$K$} &10 &0.1957 &\underline{0.2437} &\textbf{0.2749} &40.47\% \\
                & & &20 &0.3089 &\underline{0.3703} &\textbf{0.4067} &31.66\% \\
                \cline{3-8}
                & &\multirow{2}{*}{NDCG@$K$} &10 &0.0224 &\underline{0.0295} &\textbf{0.0335} &49.55\% \\
                & & &20 &0.0242 &\underline{0.0329} &\textbf{0.0379} &56.61\% \\
        \bottomrule
    \end{tabular}
    \label{tab_rec_perf}
\end{table}

\subsection{The Effectiveness of User Values for Recommendation (RQ1)}\label{sec_rq1}
We propose leveraging user values to enhance recommendation performance and generate more meaningful suggestions, as user values reflect long-term characteristics and stable, intrinsic factors that influence decision-making behavior. To effectively extract these values from user interaction histories, we introduce a multi-agent collaboration framework called ZOOM. Furthermore, we explore two strategies for integrating user values into recommender systems: direct concatenation and a contrastive learning-based auxiliary task.
Table~\ref{tab_rec_perf} reports recommendation performance in terms of HR@$K$ and NDCG@$K$ (with $K \in {10, 20}$) on two datasets. The results include three settings for the base models (i.e., MoRec and EasyRec): (a) without user value integration (``Original''), (b) with user values incorporated via direct concatenation (``DC''), and (c) with user values integrated through a contrastive learning-based auxiliary task (``CL'').

As shown in Table~\ref{tab_rec_perf}, incorporating user values, whether through the simple Direct Concatenation (DC) method or the more advanced Contrastive Learning (CL) approach, consistently improves recommendation performance, highlighting the effectiveness of leveraging user values in recommender systems.
Specifically, the DC method outperforms the Original setting in most cases across both HR and NDCG metrics. For instance, the most substantial improvement achieved by DC is observed with EasyRec on the MovieLens-1M dataset, where it delivers gains of $24.53\%$ in HR@$10$ and $35.95\%$ in NDCG@$20$, which is an impressive enhancement. However, in some scenarios, the DC method yields only comparable or slightly reduced performance. For example, in the cases of HR@$10$ for EasyRec on PENS and HR@$20$ for MoRec on MovieLens-1M, DC exhibits marginal declines of $0.97\%$ and $0.08\%$, respectively.
These declines suggest that direct concatenation may not always be optimal for integrating user values, possibly due to (1) the user value embeddings being overshadowed when concatenated with higher-dimensional user representations, and (2) potential semantic misalignment between user behavior embeddings and user value representations.
To address this, we propose a contrastive learning-based fusion strategy (CL) to more effectively incorporate user values. According to the results in Table~\ref{tab_rec_perf}, CL consistently outperforms both the Original and DC approaches across all evaluation scenarios and metrics. Notably, even in scenarios where DC performs worst, such as HR@$10$ for EasyRec on PENS and NDCG@$20$ for MoRec on MovieLens-1M, CL achieves improvements of $5.22\%$ and $6.25\%$ in HR@$10$, and $1.10\%$ and $1.15\%$ in NDCG@$20$ over the Original and DC settings, respectively.
These results underscore CL's superior ability to align and integrate user value semantics into user embeddings, ultimately leading to more effective recommendations.

Beyond the overall effectiveness demonstrated in the results, Table~\ref{tab_rec_perf} reveals an interesting phenomenon: the performance improvement of EasyRec on the MovieLens-1M dataset is particularly striking, with gains exceeding $30\%$ across all evaluation metrics. This can be attributed to the nature of EasyRec's architecture. Since EasyRec relies on collaborative filtering and generates user embeddings from textual profiles, the alignment between user embeddings and user value embeddings, which are both derived from textual content, is inherently more seamless. In contrast, MoRec employs a sequential modeling approach, where user value embeddings must be aligned with sequence-based representations, making integration more challenging.
This reasoning is further supported by comparing the relative improvements across the two base models on both datasets: EasyRec consistently shows greater performance gains than MoRec. However, on the PENS dataset, EasyRec's improvements are less pronounced than those observed on MovieLens-1M. This discrepancy can likely be explained by the differences in data sparsity between the two datasets. MovieLens-1M is denser, offering a higher average number of user interactions, which provides richer signals for inferring accurate user values. In contrast, the sparser PENS dataset offers fewer interaction data points per user, limiting the model's ability in both user value mining and user recommendation modeling.

To conclude, the results presented in Tables~\ref{tab_rec_perf} reflect the effectiveness of incorporating user values into the base recommender systems (i.e., MoRec and EasyRec), highlighting both the quality of user value extraction and the efficacy of the integration methods.

\begin{table}[]
    \centering
    \caption{Ablation Study ($F_{1}$ score $\uparrow$) for the User Values Mining Framework ZOOM. The best results are bold.}
    \begin{tabular}{lcc}
    \hline
    \textbf{} &\textbf{PENS} &\textbf{MovieLens-1M}\\
    \hline
    ZOOM & \textbf{0.6813} & \textbf{0.7282} \\
    \textit{w/o} Debate &0.2680 &0.3600 \\
    \textit{w/o} Sum &0.2068 &0.3200 \\
    Base LLM &0.1160 &0.1187 \\
    \hline
    \end{tabular}
    \label{tab_ablation}
\end{table}

\subsection{The Effectiveness of ZOOM in User Values Mining (RQ2)}
A key contribution of this work is the development of an automatic method to mine user values from users' historical interactions. In Section~\ref{sec_rq1}, we implicitly validate the effectiveness of the mined user values through the observed improvements in recommendation performance after their integration. In this section, we further evaluate the ZOOM framework's capability to accurately extract user values via a human evaluation study.

Specifically, for both datasets (PENS and MovieLens-1M), we randomly select $50$ users and compare the user values extracted by ZOOM against ground truth values provided by human evaluators. As shown in Table~\ref{tab_ablation}, ZOOM achieves $F_{1}$ scores of $0.6813$ on PENS and $0.7282$ on MovieLens-1M, which is more than six times higher than those achieved by a naive LLM-based method. These results demonstrate that ZOOM achieves a strong balance between precision and recall~\cite{hung2017computing}, and highlight its ability to generate user values that align closely with human judgment.
We further investigate the necessity of each key component of ZOOM in Section~\ref{sec_ablation}.

\subsection{Ablation Study (RQ3)} \label{sec_ablation}
To address the challenges posed by lengthy user interaction histories and the inherent hallucination problem of large language models (LLMs), we design ZOOM with two key components: a summarization module and a multi-role agent debate mechanism. In this section, we conduct an ablation study to further investigate the contribution of these components. The results are shown in Table~\ref{tab_ablation}.
We compare the $F_{1}$ scores across four settings: ZOOM, ZOOM w/o Debate, ZOOM w/o Sum, and Base LLM. In the “ZOOM” setting, user values are extracted using the fully functional ZOOM framework. In the “ZOOM w/o Debate” setting, text summarization and item truncation are retained, but the debate module is disabled. Consequently, only one evaluator (i.e., using the original chronological order) is used to extract user values, supported by a supervisor to finalize the outputs. In the “ZOOM w/o Sum” setting, the debate module and item truncation remain active, while text summarization is removed. Finally, in the “Base LLM” setting, user values are extracted directly using the original LLM, without any enhancements.

According to the results in Table~\ref{tab_ablation}, the highest $F_{1}$ scores on both datasets are achieved by the fully functional ZOOM, clearly demonstrating the effectiveness of the overall framework. In contrast, the lowest scores are observed under the Base LLM setting, which is expected due to its susceptibility to input length limitations and hallucinations. Removing the summarization module in the “ZOOM w/o Sum” setting leads to significant performance drops,  from $0.68$ to $0.20$ on PENS and from $0.72$ to $0.36$ on MovieLens-1M. This highlights the importance of summarization, as without it, the model is constrained to include only a few items (e.g., two or three) due to input length limits, resulting in a distorted view of the user's interaction history and poor value extraction.
Similarly, disabling the debate component in the “ZOOM w/o Debate” setting causes the performance to degrade by more than half, underscoring the debate module's importance in alleviating hallucinations. That is to say, the multi-agent debate enables iterative reasoning and helps verify the consistency of extracted values, thereby improving reliability.

Based on these ablation results, we conclude that both summarization and debate are essential components that significantly enhance the ZOOM framework's ability to accurately extract user values.

\begin{figure*}
    \centering
    \includegraphics[scale=0.35]{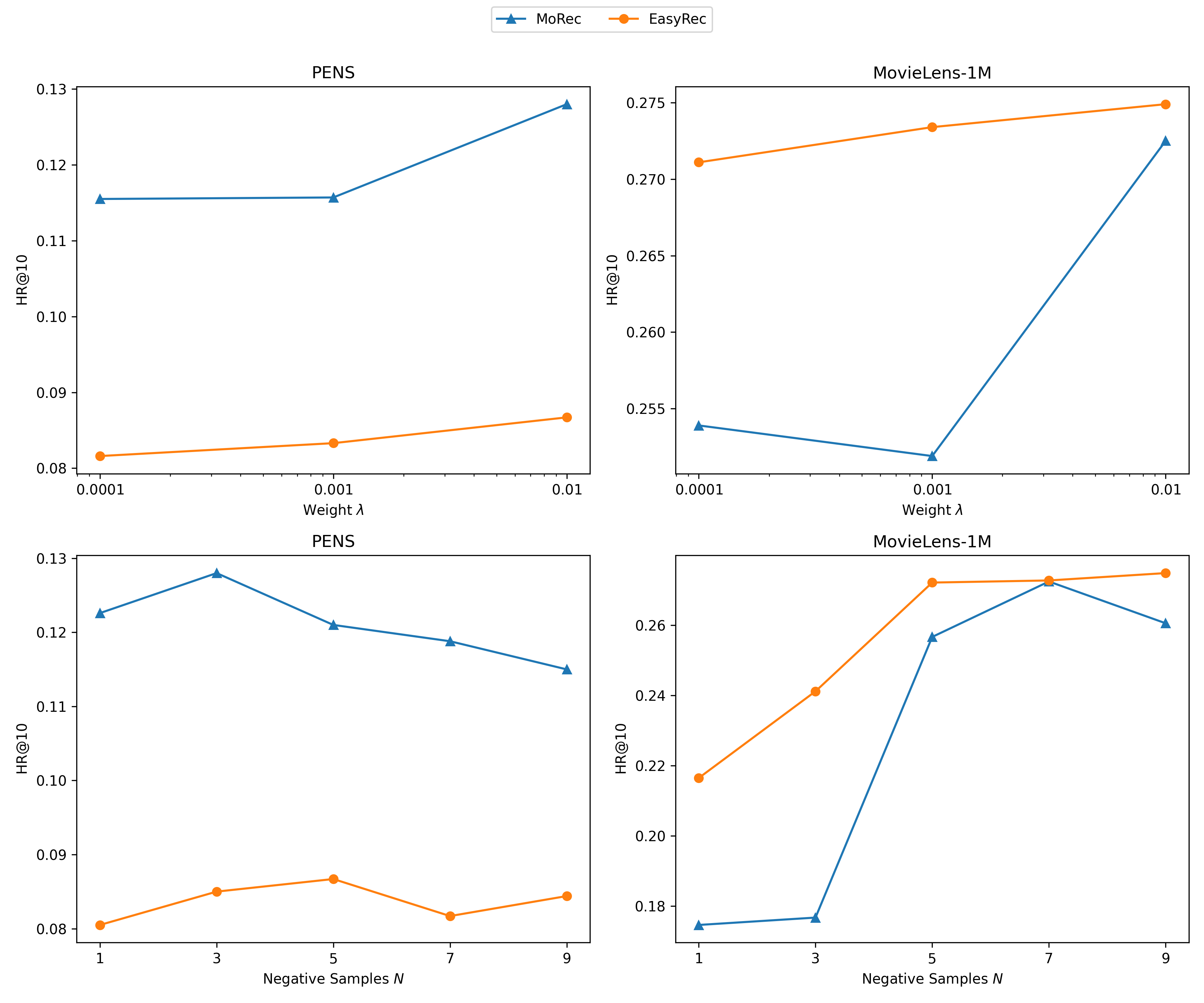}
    \caption{Analysis of Hyperparameters $\lambda$ and $N$ with respect to the HR@$10$}
    \label{fig_hyperparams}
\end{figure*}

\begin{figure*}
    \centering
    \includegraphics[scale=0.35]{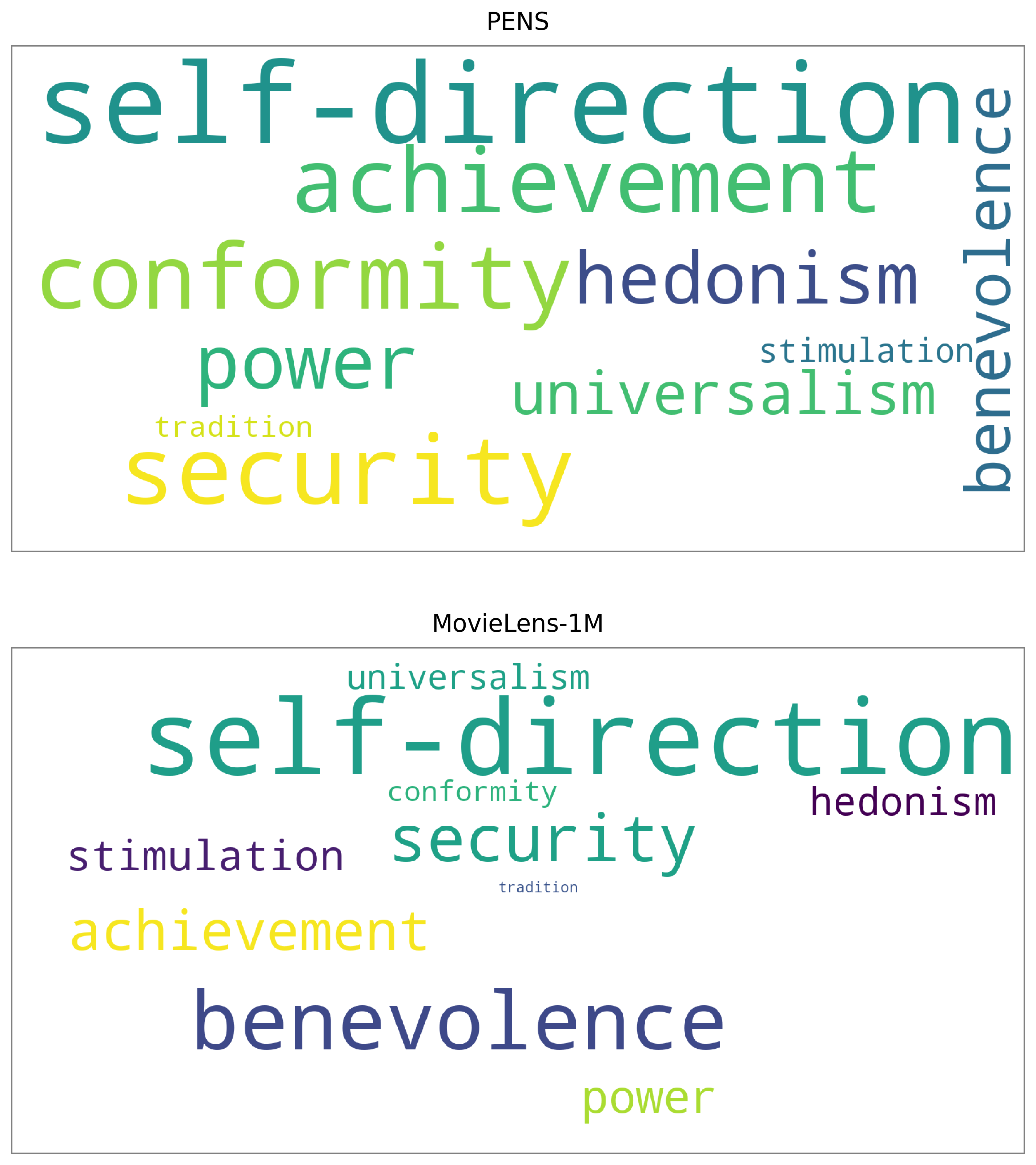}
    \caption{Word Cloud of User Values Extracted by ZOOM Based on Schwartz's Theory of Basic Values. Top: Frequency of user values in the PENS dataset. Bottom: Frequency of user values in the MovieLens-1M dataset.}
    \label{fig_uservalues_wordcloud}
\end{figure*}

\subsection{Impact of Hyperparameters (RQ4)}
To better use the mined user value, we design a contrastive learning-based auxiliary task on the integration of user values into recommender systems. 
In this part, we delve into the impact of varying hyperparameters on the integration results.
Specifically, two key hyperparameters are analyzed: $\lambda$, which controls the contribution of the contrastive learning loss, and $N$, which denotes the number of negative samples selected for each user in the contrastive learning process. 
To facilitate the analysis, when investigating the effect of a single hyperparameter, the others are held constant at their default values, with $\lambda = 0.01$ and $N = 5$. The results are reported in Figure~\ref{fig_hyperparams} with HR@$10$.

\textbf{Impact of $\lambda$.} We analyze the effect of the contrastive learning loss weight $\lambda$ using values $\left\{0.0001, 0.001, 0.01\right\}$. As observed in the top two subfigures of Figure~\ref{fig_hyperparams}, the highest HR@$10$ scores for both base models (MoRec and EasyRec) on the PENS and MovieLens-1M datasets are achieved when $\lambda = 0.01$. In other words, increasing the influence of the contrastive learning task during training (i.e., increasing $\lambda$ from $0.001$ to $0.01$) leads to improved model performance, highlighting the positive influence of incorporating contrastive learning for user value integration.
However, it is important to note that beyond a certain point, further increasing $\lambda$ may lead to diminishing returns or even degraded performance. Intuitively, as the contrastive learning component begins to dominate the optimization process, it may interfere with the core recommendation objective, ultimately reducing overall effectiveness.

\textbf{Impact of $N$.} We investigate the effect of the number of negative samples $N$ used in the contrastive learning component, considering values $\left\{1, 3, 5, 7, 9\right\}$. In our setting (see Section~\ref{sec_user_values_for_rec}), negative samples refer to user values that are unrelated to a given user, and are selected based on contrast with the user's own (positive) values.
As illustrated in the bottom two subfigures of Figure~\ref{fig_hyperparams}, the optimal value of $N$ varies across models and datasets. On the PENS dataset, the highest HR@$10$ scores are achieved when $N = 3$ for MoRec and $N = 5$ for EasyRec. In contrast, on the MovieLens-1M dataset, the optimal performance is observed at $N = 7$ for MoRec and $N = 9$ for EasyRec. These findings suggest that a larger number of negative samples is needed on the MovieLens-1M dataset to maximize recommendation accuracy.
To explain this phenomenon, we refer to the word clouds of user values extracted by ZOOM (Figure~\ref{fig_uservalues_wordcloud}). On the PENS dataset, the user value distribution is skewed toward a few dominant values, such as ``self-direction'', ``achievement'', ``conformity'', and ``security'', indicating lower diversity. In this case, even a small number of unrelated values provides sufficient contrast to effectively guide the contrastive learning process. Adding more negative samples beyond this point can introduce redundancy or noise, leading to diminished performance.
Conversely, the MovieLens-1M dataset exhibits a broader and more balanced distribution of values, including ``self-direction'', ``benevolence'', ``hedonism'', ``security'', ``power'', ``stimulation'', and ``achievement''. This higher diversity necessitates a greater number of negative samples to ensure effective contrast during learning. With fewer negative samples, the model may fail to adequately differentiate between user-specific values and unrelated ones.

\begin{figure*}
    \centering
    \includegraphics[scale=0.18]{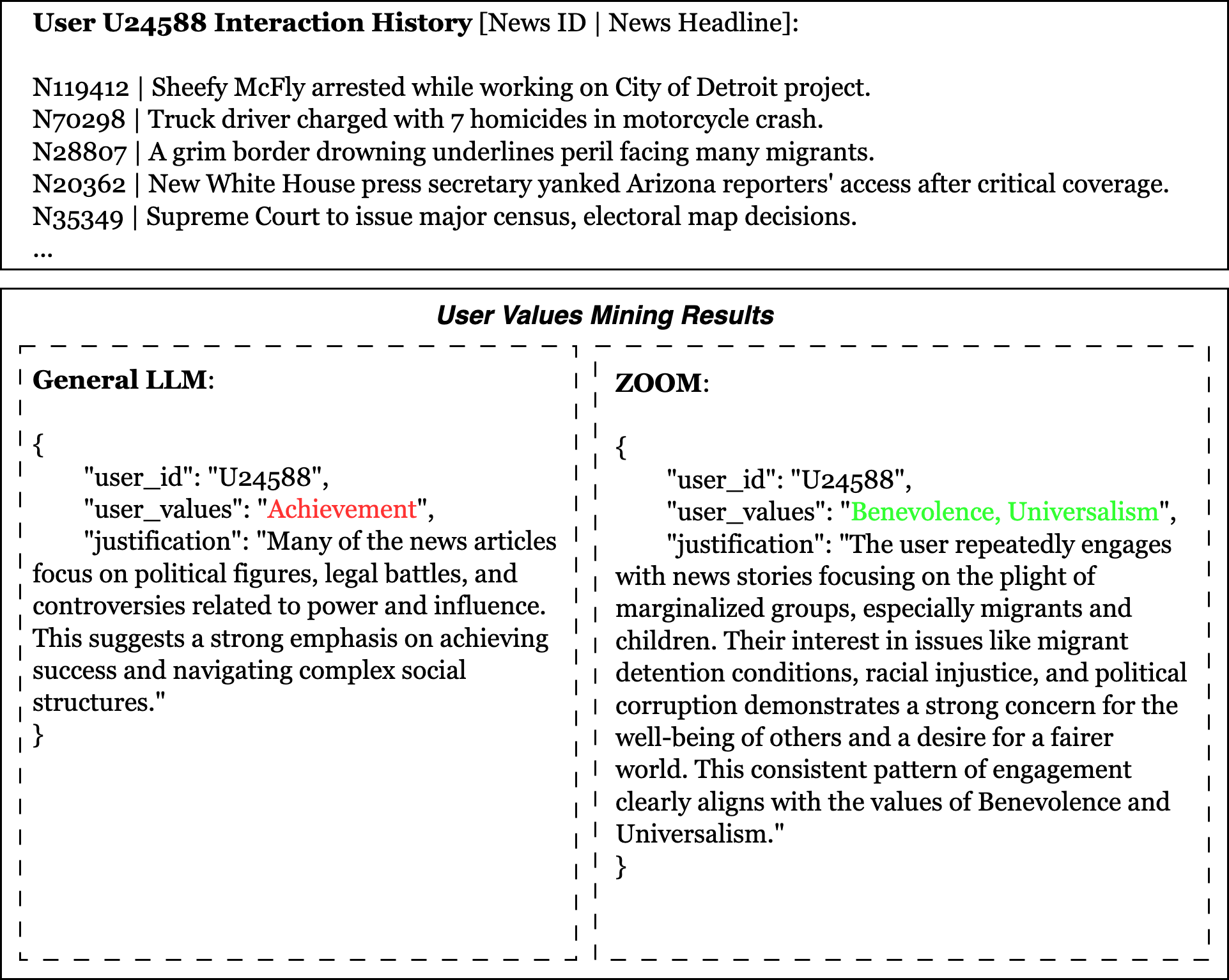}
    \caption{Case Study: Effectiveness of User Values Mining between a General LLM and ZOOM}
    \label{fig_case_study}
\end{figure*}

\subsection{Case Study} \label{sec_case_study}
We conduct a case study to demonstrate the respective performances of a general LLM and our proposed method (i.e., ZOOM) in mining user values from a given user's interaction history. We randomly select a user with ID U24588 from the PENS dataset as the study subject. For brevity, we present only a portion of the user's interaction history; specifically, for each news item in PENS, we display its ID and headline. As illustrated in Figure~\ref{fig_case_study}, the interaction history of U24588 is shown at the top, while the user value mining results provided by the general LLM and ZOOM are presented at the bottom. 

Based on the user values mining results in Figure~\ref{fig_case_study}, the general LLM identifies U24588's user value as ``Achievement''. However, considering the justification provided by the general LLM, the full interaction history of U24588 (i.e., both news headlines and content), and the discussion of user values in Section~\ref{sec_user_values}, ``Achievement'' is an inappropriate user value. In contrast, the user values ``Benevolence'' and ``Universalism'' identified by ZOOM are more appropriate, as they better align with the full meaning of the news content and are more strongly supported by the justification.

This case study highlights that, with the special design, our ZOOM can capture more appropriate and accurate user values according to user historical interactions compared to vanilla LLMs.

\section{Conclusion} \label{sec_conclusion}
In this paper, to boost recommendation performance with user values, we first propose a novel LLM agent-based framework for automatic user value discovery using users' historical interactions, namely ZOOM. It consists of two types of agents: evaluators and supervisors. The evaluators are responsible for generating user value candidates. To improve the generation performance, evaluators leverage summarization to incorporate and digest user interaction histories while generating values via multiple reasoning paths. The supervisors are designed to alleviate hallucinations by reviewing the evaluator’s output and engaging in a debate to reach a consensus on the final user values. To integrate the mined user values into recommender systems, we design a contrastive learning-based auxiliary task that enables smooth information fusion. Extensive experiments with comprehensive analysis on two real-world datasets using state-of-the-art language model-based recommender systems validate the effectiveness of the ZOOM framework in user values mining and the proposed integration strategy for incorporating user values into recommender systems.

\begin{acks}
The Australian Research Council supports this work under the streams of Future Fellowship (Grant No. FT210100624), the Discovery Project (Grant No. DP240101108), and the Linkage Project (Grant No. LP230200892).
\end{acks}

\bibliographystyle{ACM-Reference-Format}
\bibliography{sample-base}


\begin{thebibliography}{68}

%%% the Web address.

\ifx \showCODEN    \undefined \def \showCODEN     #1{\unskip}     \fi
\ifx \showDOI      \undefined \def \showDOI       #1{#1}\fi
\ifx \showISBNx    \undefined \def \showISBNx     #1{\unskip}     \fi
\ifx \showISBNxiii \undefined \def \showISBNxiii  #1{\unskip}     \fi
\ifx \showISSN     \undefined \def \showISSN      #1{\unskip}     \fi
\ifx \showLCCN     \undefined \def \showLCCN      #1{\unskip}     \fi
\ifx \shownote     \undefined \def \shownote      #1{#1}          \fi
\ifx \showarticletitle \undefined \def \showarticletitle #1{#1}   \fi
\ifx \showURL      \undefined \def \showURL       {\relax}        \fi
% invisible to TeX
\providecommand\bibfield[2]{#2}
\providecommand\bibinfo[2]{#2}
\providecommand\natexlab[1]{#1}
\providecommand\showeprint[2][]{arXiv:#2}

\bibitem[Ao et~al\mbox{.}(2021)]%
        {ao2021pens}
\bibfield{author}{\bibinfo{person}{Xiang Ao}, \bibinfo{person}{Xiting Wang}, \bibinfo{person}{Ling Luo}, \bibinfo{person}{Ying Qiao}, \bibinfo{person}{Qing He}, {and} \bibinfo{person}{Xing Xie}.} \bibinfo{year}{2021}\natexlab{}.
\newblock \showarticletitle{PENS: A dataset and generic framework for personalized news headline generation}. In \bibinfo{booktitle}{\emph{Proceedings of the 59th Annual Meeting of the Association for Computational Linguistics and the 11th International Joint Conference on Natural Language Processing (Volume 1: Long Papers)}}. \bibinfo{pages}{82--92}.
\newblock


\bibitem[Brandt(2017)]%
        {brandt2017predicting}
\bibfield{author}{\bibinfo{person}{Mark~J Brandt}.} \bibinfo{year}{2017}\natexlab{}.
\newblock \showarticletitle{Predicting ideological prejudice}.
\newblock \bibinfo{journal}{\emph{Psychological Science}} \bibinfo{volume}{28}, \bibinfo{number}{6} (\bibinfo{year}{2017}), \bibinfo{pages}{713--722}.
\newblock


\bibitem[Burns et~al\mbox{.}(2022)]%
        {burns2022discovering}
\bibfield{author}{\bibinfo{person}{Collin Burns}, \bibinfo{person}{Haotian Ye}, \bibinfo{person}{Dan Klein}, {and} \bibinfo{person}{Jacob Steinhardt}.} \bibinfo{year}{2022}\natexlab{}.
\newblock \showarticletitle{Discovering latent knowledge in language models without supervision}.
\newblock \bibinfo{journal}{\emph{arXiv preprint arXiv:2212.03827}} (\bibinfo{year}{2022}).
\newblock


\bibitem[Chen et~al\mbox{.}(2024)]%
        {chen2024adversarial}
\bibfield{author}{\bibinfo{person}{Lijian Chen}, \bibinfo{person}{Wei Yuan}, \bibinfo{person}{Tong Chen}, \bibinfo{person}{Guanhua Ye}, \bibinfo{person}{Nguyen Quoc~Viet Hung}, {and} \bibinfo{person}{Hongzhi Yin}.} \bibinfo{year}{2024}\natexlab{}.
\newblock \showarticletitle{Adversarial item promotion on visually-aware recommender systems by guided diffusion}.
\newblock \bibinfo{journal}{\emph{ACM Transactions on Information Systems}} \bibinfo{volume}{42}, \bibinfo{number}{6} (\bibinfo{year}{2024}), \bibinfo{pages}{1--26}.
\newblock


\bibitem[Chowdhery et~al\mbox{.}(2023)]%
        {chowdhery2023palm}
\bibfield{author}{\bibinfo{person}{Aakanksha Chowdhery}, \bibinfo{person}{Sharan Narang}, \bibinfo{person}{Jacob Devlin}, \bibinfo{person}{Maarten Bosma}, \bibinfo{person}{Gaurav Mishra}, \bibinfo{person}{Adam Roberts}, \bibinfo{person}{Paul Barham}, \bibinfo{person}{Hyung~Won Chung}, \bibinfo{person}{Charles Sutton}, \bibinfo{person}{Sebastian Gehrmann}, {et~al\mbox{.}}} \bibinfo{year}{2023}\natexlab{}.
\newblock \showarticletitle{Palm: Scaling language modeling with pathways}.
\newblock \bibinfo{journal}{\emph{Journal of Machine Learning Research}} \bibinfo{volume}{24}, \bibinfo{number}{240} (\bibinfo{year}{2023}), \bibinfo{pages}{1--113}.
\newblock


\bibitem[Christiano et~al\mbox{.}(2017)]%
        {christiano2017deep}
\bibfield{author}{\bibinfo{person}{Paul~F Christiano}, \bibinfo{person}{Jan Leike}, \bibinfo{person}{Tom Brown}, \bibinfo{person}{Miljan Martic}, \bibinfo{person}{Shane Legg}, {and} \bibinfo{person}{Dario Amodei}.} \bibinfo{year}{2017}\natexlab{}.
\newblock \showarticletitle{Deep reinforcement learning from human preferences}.
\newblock \bibinfo{journal}{\emph{Advances in neural information processing systems}}  \bibinfo{volume}{30} (\bibinfo{year}{2017}).
\newblock


\bibitem[Chung et~al\mbox{.}(2024)]%
        {chung2024scaling}
\bibfield{author}{\bibinfo{person}{Hyung~Won Chung}, \bibinfo{person}{Le Hou}, \bibinfo{person}{Shayne Longpre}, \bibinfo{person}{Barret Zoph}, \bibinfo{person}{Yi Tay}, \bibinfo{person}{William Fedus}, \bibinfo{person}{Yunxuan Li}, \bibinfo{person}{Xuezhi Wang}, \bibinfo{person}{Mostafa Dehghani}, \bibinfo{person}{Siddhartha Brahma}, {et~al\mbox{.}}} \bibinfo{year}{2024}\natexlab{}.
\newblock \showarticletitle{Scaling instruction-finetuned language models}.
\newblock \bibinfo{journal}{\emph{Journal of Machine Learning Research}} \bibinfo{volume}{25}, \bibinfo{number}{70} (\bibinfo{year}{2024}), \bibinfo{pages}{1--53}.
\newblock


\bibitem[Dosovitskiy et~al\mbox{.}(2020)]%
        {dosovitskiy2020image}
\bibfield{author}{\bibinfo{person}{Alexey Dosovitskiy}, \bibinfo{person}{Lucas Beyer}, \bibinfo{person}{Alexander Kolesnikov}, \bibinfo{person}{Dirk Weissenborn}, \bibinfo{person}{Xiaohua Zhai}, \bibinfo{person}{Thomas Unterthiner}, \bibinfo{person}{Mostafa Dehghani}, \bibinfo{person}{Matthias Minderer}, \bibinfo{person}{Georg Heigold}, \bibinfo{person}{Sylvain Gelly}, {et~al\mbox{.}}} \bibinfo{year}{2020}\natexlab{}.
\newblock \showarticletitle{An image is worth 16x16 words: Transformers for image recognition at scale}.
\newblock \bibinfo{journal}{\emph{arXiv preprint arXiv:2010.11929}} (\bibinfo{year}{2020}).
\newblock


\bibitem[Du et~al\mbox{.}(2023)]%
        {du2023improving}
\bibfield{author}{\bibinfo{person}{Yilun Du}, \bibinfo{person}{Shuang Li}, \bibinfo{person}{Antonio Torralba}, \bibinfo{person}{Joshua~B Tenenbaum}, {and} \bibinfo{person}{Igor Mordatch}.} \bibinfo{year}{2023}\natexlab{}.
\newblock \showarticletitle{Improving factuality and reasoning in language models through multiagent debate}. In \bibinfo{booktitle}{\emph{Forty-first International Conference on Machine Learning}}.
\newblock


\bibitem[Dubey et~al\mbox{.}(2024)]%
        {dubey2024llama}
\bibfield{author}{\bibinfo{person}{Abhimanyu Dubey}, \bibinfo{person}{Abhinav Jauhri}, \bibinfo{person}{Abhinav Pandey}, \bibinfo{person}{Abhishek Kadian}, \bibinfo{person}{Ahmad Al-Dahle}, \bibinfo{person}{Aiesha Letman}, \bibinfo{person}{Akhil Mathur}, \bibinfo{person}{Alan Schelten}, \bibinfo{person}{Amy Yang}, \bibinfo{person}{Angela Fan}, {et~al\mbox{.}}} \bibinfo{year}{2024}\natexlab{}.
\newblock \showarticletitle{The llama 3 herd of models}.
\newblock \bibinfo{journal}{\emph{arXiv preprint arXiv:2407.21783}} (\bibinfo{year}{2024}).
\newblock


\bibitem[Fan et~al\mbox{.}(2018)]%
        {fan2018hierarchical}
\bibfield{author}{\bibinfo{person}{Angela Fan}, \bibinfo{person}{Mike Lewis}, {and} \bibinfo{person}{Yann Dauphin}.} \bibinfo{year}{2018}\natexlab{}.
\newblock \showarticletitle{Hierarchical neural story generation}.
\newblock \bibinfo{journal}{\emph{arXiv preprint arXiv:1805.04833}} (\bibinfo{year}{2018}).
\newblock


\bibitem[Ficler and Goldberg(2017)]%
        {ficler2017controlling}
\bibfield{author}{\bibinfo{person}{Jessica Ficler} {and} \bibinfo{person}{Yoav Goldberg}.} \bibinfo{year}{2017}\natexlab{}.
\newblock \showarticletitle{Controlling linguistic style aspects in neural language generation}.
\newblock \bibinfo{journal}{\emph{arXiv preprint arXiv:1707.02633}} (\bibinfo{year}{2017}).
\newblock


\bibitem[Harper and Konstan(2015)]%
        {harper2015movielens}
\bibfield{author}{\bibinfo{person}{F~Maxwell Harper} {and} \bibinfo{person}{Joseph~A Konstan}.} \bibinfo{year}{2015}\natexlab{}.
\newblock \showarticletitle{The movielens datasets: History and context}.
\newblock \bibinfo{journal}{\emph{Acm transactions on interactive intelligent systems (tiis)}} \bibinfo{volume}{5}, \bibinfo{number}{4} (\bibinfo{year}{2015}), \bibinfo{pages}{1--19}.
\newblock


\bibitem[He and McAuley(2016)]%
        {he2016vbpr}
\bibfield{author}{\bibinfo{person}{Ruining He} {and} \bibinfo{person}{Julian McAuley}.} \bibinfo{year}{2016}\natexlab{}.
\newblock \showarticletitle{VBPR: visual bayesian personalized ranking from implicit feedback}. In \bibinfo{booktitle}{\emph{Proceedings of the AAAI conference on artificial intelligence}}, Vol.~\bibinfo{volume}{30}.
\newblock


\bibitem[He et~al\mbox{.}(2017)]%
        {he2017neural}
\bibfield{author}{\bibinfo{person}{Xiangnan He}, \bibinfo{person}{Lizi Liao}, \bibinfo{person}{Hanwang Zhang}, \bibinfo{person}{Liqiang Nie}, \bibinfo{person}{Xia Hu}, {and} \bibinfo{person}{Tat-Seng Chua}.} \bibinfo{year}{2017}\natexlab{}.
\newblock \showarticletitle{Neural collaborative filtering}. In \bibinfo{booktitle}{\emph{Proceedings of the 26th international conference on world wide web}}. \bibinfo{pages}{173--182}.
\newblock


\bibitem[Hidasi et~al\mbox{.}(2015)]%
        {hidasi2015session}
\bibfield{author}{\bibinfo{person}{Bal{\'a}zs Hidasi}, \bibinfo{person}{Alexandros Karatzoglou}, \bibinfo{person}{Linas Baltrunas}, {and} \bibinfo{person}{Domonkos Tikk}.} \bibinfo{year}{2015}\natexlab{}.
\newblock \showarticletitle{Session-based recommendations with recurrent neural networks}.
\newblock \bibinfo{journal}{\emph{arXiv preprint arXiv:1511.06939}} (\bibinfo{year}{2015}).
\newblock


\bibitem[Holtzman et~al\mbox{.}(2019)]%
        {holtzman2019curious}
\bibfield{author}{\bibinfo{person}{Ari Holtzman}, \bibinfo{person}{Jan Buys}, \bibinfo{person}{Li Du}, \bibinfo{person}{Maxwell Forbes}, {and} \bibinfo{person}{Yejin Choi}.} \bibinfo{year}{2019}\natexlab{}.
\newblock \showarticletitle{The curious case of neural text degeneration}.
\newblock \bibinfo{journal}{\emph{arXiv preprint arXiv:1904.09751}} (\bibinfo{year}{2019}).
\newblock


\bibitem[Holtzman et~al\mbox{.}(2018)]%
        {holtzman2018learning}
\bibfield{author}{\bibinfo{person}{Ari Holtzman}, \bibinfo{person}{Jan Buys}, \bibinfo{person}{Maxwell Forbes}, \bibinfo{person}{Antoine Bosselut}, \bibinfo{person}{David Golub}, {and} \bibinfo{person}{Yejin Choi}.} \bibinfo{year}{2018}\natexlab{}.
\newblock \showarticletitle{Learning to write with cooperative discriminators}.
\newblock \bibinfo{journal}{\emph{arXiv preprint arXiv:1805.06087}} (\bibinfo{year}{2018}).
\newblock


\bibitem[Huang et~al\mbox{.}(2023)]%
        {huang2023survey}
\bibfield{author}{\bibinfo{person}{Lei Huang}, \bibinfo{person}{Weijiang Yu}, \bibinfo{person}{Weitao Ma}, \bibinfo{person}{Weihong Zhong}, \bibinfo{person}{Zhangyin Feng}, \bibinfo{person}{Haotian Wang}, \bibinfo{person}{Qianglong Chen}, \bibinfo{person}{Weihua Peng}, \bibinfo{person}{Xiaocheng Feng}, \bibinfo{person}{Bing Qin}, {et~al\mbox{.}}} \bibinfo{year}{2023}\natexlab{}.
\newblock \showarticletitle{A survey on hallucination in large language models: Principles, taxonomy, challenges, and open questions}.
\newblock \bibinfo{journal}{\emph{ACM Transactions on Information Systems}} (\bibinfo{year}{2023}).
\newblock


\bibitem[Hung et~al\mbox{.}(2017)]%
        {hung2017computing}
\bibfield{author}{\bibinfo{person}{Nguyen Quoc~Viet Hung}, \bibinfo{person}{Huynh~Huu Viet}, \bibinfo{person}{Nguyen~Thanh Tam}, \bibinfo{person}{Matthias Weidlich}, \bibinfo{person}{Hongzhi Yin}, {and} \bibinfo{person}{Xiaofang Zhou}.} \bibinfo{year}{2017}\natexlab{}.
\newblock \showarticletitle{Computing crowd consensus with partial agreement}.
\newblock \bibinfo{journal}{\emph{IEEE Transactions on Knowledge and Data Engineering}} \bibinfo{volume}{30}, \bibinfo{number}{1} (\bibinfo{year}{2017}), \bibinfo{pages}{1--14}.
\newblock


\bibitem[Jaskolka et~al\mbox{.}(1985)]%
        {jaskolka1985measuring}
\bibfield{author}{\bibinfo{person}{Gabriel Jaskolka}, \bibinfo{person}{Janice~M Beyer}, {and} \bibinfo{person}{Harrison~M Trice}.} \bibinfo{year}{1985}\natexlab{}.
\newblock \showarticletitle{Measuring and predicting managerial success}.
\newblock \bibinfo{journal}{\emph{Journal of vocational behavior}} \bibinfo{volume}{26}, \bibinfo{number}{2} (\bibinfo{year}{1985}), \bibinfo{pages}{189--205}.
\newblock


\bibitem[Ji et~al\mbox{.}(2023)]%
        {ji2023survey}
\bibfield{author}{\bibinfo{person}{Ziwei Ji}, \bibinfo{person}{Nayeon Lee}, \bibinfo{person}{Rita Frieske}, \bibinfo{person}{Tiezheng Yu}, \bibinfo{person}{Dan Su}, \bibinfo{person}{Yan Xu}, \bibinfo{person}{Etsuko Ishii}, \bibinfo{person}{Ye~Jin Bang}, \bibinfo{person}{Andrea Madotto}, {and} \bibinfo{person}{Pascale Fung}.} \bibinfo{year}{2023}\natexlab{}.
\newblock \showarticletitle{Survey of hallucination in natural language generation}.
\newblock \bibinfo{journal}{\emph{Comput. Surveys}} \bibinfo{volume}{55}, \bibinfo{number}{12} (\bibinfo{year}{2023}), \bibinfo{pages}{1--38}.
\newblock


\bibitem[Kang and McAuley(2018)]%
        {kang2018self}
\bibfield{author}{\bibinfo{person}{Wang-Cheng Kang} {and} \bibinfo{person}{Julian McAuley}.} \bibinfo{year}{2018}\natexlab{}.
\newblock \showarticletitle{Self-attentive sequential recommendation}. In \bibinfo{booktitle}{\emph{2018 IEEE international conference on data mining (ICDM)}}. IEEE, \bibinfo{pages}{197--206}.
\newblock


\bibitem[Koren et~al\mbox{.}(2009)]%
        {koren2009matrix}
\bibfield{author}{\bibinfo{person}{Yehuda Koren}, \bibinfo{person}{Robert Bell}, {and} \bibinfo{person}{Chris Volinsky}.} \bibinfo{year}{2009}\natexlab{}.
\newblock \showarticletitle{Matrix factorization techniques for recommender systems}.
\newblock \bibinfo{journal}{\emph{Computer}} \bibinfo{volume}{42}, \bibinfo{number}{8} (\bibinfo{year}{2009}), \bibinfo{pages}{30--37}.
\newblock


\bibitem[Krichene and Rendle(2020)]%
        {krichene2020sampled}
\bibfield{author}{\bibinfo{person}{Walid Krichene} {and} \bibinfo{person}{Steffen Rendle}.} \bibinfo{year}{2020}\natexlab{}.
\newblock \showarticletitle{On sampled metrics for item recommendation}. In \bibinfo{booktitle}{\emph{Proceedings of the 26th ACM SIGKDD international conference on knowledge discovery \& data mining}}. \bibinfo{pages}{1748--1757}.
\newblock


\bibitem[Kwon et~al\mbox{.}(2023)]%
        {kwon2023efficient}
\bibfield{author}{\bibinfo{person}{Woosuk Kwon}, \bibinfo{person}{Zhuohan Li}, \bibinfo{person}{Siyuan Zhuang}, \bibinfo{person}{Ying Sheng}, \bibinfo{person}{Lianmin Zheng}, \bibinfo{person}{Cody~Hao Yu}, \bibinfo{person}{Joseph~E. Gonzalez}, \bibinfo{person}{Hao Zhang}, {and} \bibinfo{person}{Ion Stoica}.} \bibinfo{year}{2023}\natexlab{}.
\newblock \showarticletitle{Efficient Memory Management for Large Language Model Serving with PagedAttention}. In \bibinfo{booktitle}{\emph{Proceedings of the ACM SIGOPS 29th Symposium on Operating Systems Principles}}.
\newblock


\bibitem[Li et~al\mbox{.}(2023)]%
        {li2023adapting}
\bibfield{author}{\bibinfo{person}{Qingyao Li}, \bibinfo{person}{Lingyue Fu}, \bibinfo{person}{Weiming Zhang}, \bibinfo{person}{Xianyu Chen}, \bibinfo{person}{Jingwei Yu}, \bibinfo{person}{Wei Xia}, \bibinfo{person}{Weinan Zhang}, \bibinfo{person}{Ruiming Tang}, {and} \bibinfo{person}{Yong Yu}.} \bibinfo{year}{2023}\natexlab{}.
\newblock \showarticletitle{Adapting large language models for education: Foundational capabilities, potentials, and challenges}.
\newblock \bibinfo{journal}{\emph{arXiv preprint arXiv:2401.08664}} (\bibinfo{year}{2023}).
\newblock


\bibitem[Li et~al\mbox{.}(2024)]%
        {li2024multi}
\bibfield{author}{\bibinfo{person}{Youhua Li}, \bibinfo{person}{Hanwen Du}, \bibinfo{person}{Yongxin Ni}, \bibinfo{person}{Pengpeng Zhao}, \bibinfo{person}{Qi Guo}, \bibinfo{person}{Fajie Yuan}, {and} \bibinfo{person}{Xiaofang Zhou}.} \bibinfo{year}{2024}\natexlab{}.
\newblock \showarticletitle{Multi-modality is all you need for transferable recommender systems}. In \bibinfo{booktitle}{\emph{2024 IEEE 40th International Conference on Data Engineering (ICDE)}}. IEEE, \bibinfo{pages}{5008--5021}.
\newblock


\bibitem[Liu et~al\mbox{.}(2019)]%
        {liu2019roberta}
\bibfield{author}{\bibinfo{person}{Yinhan Liu}, \bibinfo{person}{Myle Ott}, \bibinfo{person}{Naman Goyal}, \bibinfo{person}{Jingfei Du}, \bibinfo{person}{Mandar Joshi}, \bibinfo{person}{Danqi Chen}, \bibinfo{person}{Omer Levy}, \bibinfo{person}{Mike Lewis}, \bibinfo{person}{Luke Zettlemoyer}, {and} \bibinfo{person}{Veselin Stoyanov}.} \bibinfo{year}{2019}\natexlab{}.
\newblock \showarticletitle{Roberta: A robustly optimized bert pretraining approach}.
\newblock \bibinfo{journal}{\emph{arXiv preprint arXiv:1907.11692}} (\bibinfo{year}{2019}).
\newblock


\bibitem[Magnusson et~al\mbox{.}(2019)]%
        {magnusson2019bayesian}
\bibfield{author}{\bibinfo{person}{M{\aa}ns Magnusson}, \bibinfo{person}{Michael Andersen}, \bibinfo{person}{Johan Jonasson}, {and} \bibinfo{person}{Aki Vehtari}.} \bibinfo{year}{2019}\natexlab{}.
\newblock \showarticletitle{Bayesian leave-one-out cross-validation for large data}. In \bibinfo{booktitle}{\emph{International Conference on Machine Learning}}. PMLR, \bibinfo{pages}{4244--4253}.
\newblock


\bibitem[Mann et~al\mbox{.}(2020)]%
        {mann2020language}
\bibfield{author}{\bibinfo{person}{Ben Mann}, \bibinfo{person}{N Ryder}, \bibinfo{person}{M Subbiah}, \bibinfo{person}{J Kaplan}, \bibinfo{person}{P Dhariwal}, \bibinfo{person}{A Neelakantan}, \bibinfo{person}{P Shyam}, \bibinfo{person}{G Sastry}, \bibinfo{person}{A Askell}, \bibinfo{person}{S Agarwal}, {et~al\mbox{.}}} \bibinfo{year}{2020}\natexlab{}.
\newblock \showarticletitle{Language models are few-shot learners}.
\newblock \bibinfo{journal}{\emph{arXiv preprint arXiv:2005.14165}}  \bibinfo{volume}{1} (\bibinfo{year}{2020}), \bibinfo{pages}{3}.
\newblock


\bibitem[Maynez et~al\mbox{.}(2020)]%
        {maynez2020faithfulness}
\bibfield{author}{\bibinfo{person}{Joshua Maynez}, \bibinfo{person}{Shashi Narayan}, \bibinfo{person}{Bernd Bohnet}, {and} \bibinfo{person}{Ryan McDonald}.} \bibinfo{year}{2020}\natexlab{}.
\newblock \showarticletitle{On faithfulness and factuality in abstractive summarization}.
\newblock \bibinfo{journal}{\emph{arXiv preprint arXiv:2005.00661}} (\bibinfo{year}{2020}).
\newblock


\bibitem[Ouyang et~al\mbox{.}(2022)]%
        {ouyang2022training}
\bibfield{author}{\bibinfo{person}{Long Ouyang}, \bibinfo{person}{Jeffrey Wu}, \bibinfo{person}{Xu Jiang}, \bibinfo{person}{Diogo Almeida}, \bibinfo{person}{Carroll Wainwright}, \bibinfo{person}{Pamela Mishkin}, \bibinfo{person}{Chong Zhang}, \bibinfo{person}{Sandhini Agarwal}, \bibinfo{person}{Katarina Slama}, \bibinfo{person}{Alex Ray}, {et~al\mbox{.}}} \bibinfo{year}{2022}\natexlab{}.
\newblock \showarticletitle{Training language models to follow instructions with human feedback}.
\newblock \bibinfo{journal}{\emph{Advances in neural information processing systems}}  \bibinfo{volume}{35} (\bibinfo{year}{2022}), \bibinfo{pages}{27730--27744}.
\newblock


\bibitem[Paszke et~al\mbox{.}(2019)]%
        {paszke2019pytorch}
\bibfield{author}{\bibinfo{person}{Adam Paszke}, \bibinfo{person}{Sam Gross}, \bibinfo{person}{Francisco Massa}, \bibinfo{person}{Adam Lerer}, \bibinfo{person}{James Bradbury}, \bibinfo{person}{Gregory Chanan}, \bibinfo{person}{Trevor Killeen}, \bibinfo{person}{Zeming Lin}, \bibinfo{person}{Natalia Gimelshein}, \bibinfo{person}{Luca Antiga}, {et~al\mbox{.}}} \bibinfo{year}{2019}\natexlab{}.
\newblock \showarticletitle{Pytorch: An imperative style, high-performance deep learning library}.
\newblock \bibinfo{journal}{\emph{Advances in neural information processing systems}}  \bibinfo{volume}{32} (\bibinfo{year}{2019}).
\newblock


\bibitem[Peng et~al\mbox{.}(2023)]%
        {peng2023instruction}
\bibfield{author}{\bibinfo{person}{Baolin Peng}, \bibinfo{person}{Chunyuan Li}, \bibinfo{person}{Pengcheng He}, \bibinfo{person}{Michel Galley}, {and} \bibinfo{person}{Jianfeng Gao}.} \bibinfo{year}{2023}\natexlab{}.
\newblock \showarticletitle{Instruction tuning with gpt-4}.
\newblock \bibinfo{journal}{\emph{arXiv preprint arXiv:2304.03277}} (\bibinfo{year}{2023}).
\newblock


\bibitem[Ren and Huang(2024)]%
        {ren2024easyrec}
\bibfield{author}{\bibinfo{person}{Xubin Ren} {and} \bibinfo{person}{Chao Huang}.} \bibinfo{year}{2024}\natexlab{}.
\newblock \showarticletitle{EasyRec: Simple yet Effective Language Models for Recommendation}.
\newblock \bibinfo{journal}{\emph{arXiv preprint arXiv:2408.08821}} (\bibinfo{year}{2024}).
\newblock


\bibitem[Rendle et~al\mbox{.}(2012)]%
        {rendle2012bpr}
\bibfield{author}{\bibinfo{person}{Steffen Rendle}, \bibinfo{person}{Christoph Freudenthaler}, \bibinfo{person}{Zeno Gantner}, {and} \bibinfo{person}{Lars Schmidt-Thieme}.} \bibinfo{year}{2012}\natexlab{}.
\newblock \showarticletitle{BPR: Bayesian personalized ranking from implicit feedback}.
\newblock \bibinfo{journal}{\emph{arXiv preprint arXiv:1205.2618}} (\bibinfo{year}{2012}).
\newblock


\bibitem[Rendle et~al\mbox{.}(2010)]%
        {rendle2010factorizing}
\bibfield{author}{\bibinfo{person}{Steffen Rendle}, \bibinfo{person}{Christoph Freudenthaler}, {and} \bibinfo{person}{Lars Schmidt-Thieme}.} \bibinfo{year}{2010}\natexlab{}.
\newblock \showarticletitle{Factorizing personalized markov chains for next-basket recommendation}. In \bibinfo{booktitle}{\emph{Proceedings of the 19th international conference on World wide web}}. \bibinfo{pages}{811--820}.
\newblock


\bibitem[Sarwar et~al\mbox{.}(2001)]%
        {sarwar2001item}
\bibfield{author}{\bibinfo{person}{Badrul Sarwar}, \bibinfo{person}{George Karypis}, \bibinfo{person}{Joseph Konstan}, {and} \bibinfo{person}{John Riedl}.} \bibinfo{year}{2001}\natexlab{}.
\newblock \showarticletitle{Item-based collaborative filtering recommendation algorithms}. In \bibinfo{booktitle}{\emph{Proceedings of the 10th international conference on World Wide Web}}. \bibinfo{pages}{285--295}.
\newblock


\bibitem[Schafer et~al\mbox{.}(1999)]%
        {schafer1999recommender}
\bibfield{author}{\bibinfo{person}{J~Ben Schafer}, \bibinfo{person}{Joseph Konstan}, {and} \bibinfo{person}{John Riedl}.} \bibinfo{year}{1999}\natexlab{}.
\newblock \showarticletitle{Recommender systems in e-commerce}. In \bibinfo{booktitle}{\emph{Proceedings of the 1st ACM conference on Electronic commerce}}. \bibinfo{pages}{158--166}.
\newblock


\bibitem[Schwartz(2012)]%
        {schwartz2012overview}
\bibfield{author}{\bibinfo{person}{Shalom~H Schwartz}.} \bibinfo{year}{2012}\natexlab{}.
\newblock \showarticletitle{An overview of the Schwartz theory of basic values}.
\newblock \bibinfo{journal}{\emph{Online readings in Psychology and Culture}} \bibinfo{volume}{2}, \bibinfo{number}{1} (\bibinfo{year}{2012}), \bibinfo{pages}{11}.
\newblock


\bibitem[Stiennon et~al\mbox{.}(2020)]%
        {stiennon2020learning}
\bibfield{author}{\bibinfo{person}{Nisan Stiennon}, \bibinfo{person}{Long Ouyang}, \bibinfo{person}{Jeffrey Wu}, \bibinfo{person}{Daniel Ziegler}, \bibinfo{person}{Ryan Lowe}, \bibinfo{person}{Chelsea Voss}, \bibinfo{person}{Alec Radford}, \bibinfo{person}{Dario Amodei}, {and} \bibinfo{person}{Paul~F Christiano}.} \bibinfo{year}{2020}\natexlab{}.
\newblock \showarticletitle{Learning to summarize with human feedback}.
\newblock \bibinfo{journal}{\emph{Advances in neural information processing systems}}  \bibinfo{volume}{33} (\bibinfo{year}{2020}), \bibinfo{pages}{3008--3021}.
\newblock


\bibitem[Tang et~al\mbox{.}(2023)]%
        {tang2023does}
\bibfield{author}{\bibinfo{person}{Ruixiang Tang}, \bibinfo{person}{Xiaotian Han}, \bibinfo{person}{Xiaoqian Jiang}, {and} \bibinfo{person}{Xia Hu}.} \bibinfo{year}{2023}\natexlab{}.
\newblock \showarticletitle{Does synthetic data generation of llms help clinical text mining?}
\newblock \bibinfo{journal}{\emph{arXiv preprint arXiv:2303.04360}} (\bibinfo{year}{2023}).
\newblock


\bibitem[Tay et~al\mbox{.}(2018)]%
        {tay2018latent}
\bibfield{author}{\bibinfo{person}{Yi Tay}, \bibinfo{person}{Luu Anh~Tuan}, {and} \bibinfo{person}{Siu~Cheung Hui}.} \bibinfo{year}{2018}\natexlab{}.
\newblock \showarticletitle{Latent relational metric learning via memory-based attention for collaborative ranking}. In \bibinfo{booktitle}{\emph{Proceedings of the 2018 world wide web conference}}. \bibinfo{pages}{729--739}.
\newblock


\bibitem[Team et~al\mbox{.}(2024)]%
        {team2024gemma}
\bibfield{author}{\bibinfo{person}{Gemma Team}, \bibinfo{person}{Morgane Riviere}, \bibinfo{person}{Shreya Pathak}, \bibinfo{person}{Pier~Giuseppe Sessa}, \bibinfo{person}{Cassidy Hardin}, \bibinfo{person}{Surya Bhupatiraju}, \bibinfo{person}{L{\'e}onard Hussenot}, \bibinfo{person}{Thomas Mesnard}, \bibinfo{person}{Bobak Shahriari}, \bibinfo{person}{Alexandre Ram{\'e}}, {et~al\mbox{.}}} \bibinfo{year}{2024}\natexlab{}.
\newblock \showarticletitle{Gemma 2: Improving open language models at a practical size}.
\newblock \bibinfo{journal}{\emph{arXiv preprint arXiv:2408.00118}} (\bibinfo{year}{2024}).
\newblock


\bibitem[Touvron et~al\mbox{.}(2023)]%
        {touvron2023llama}
\bibfield{author}{\bibinfo{person}{Hugo Touvron}, \bibinfo{person}{Louis Martin}, \bibinfo{person}{Kevin Stone}, \bibinfo{person}{Peter Albert}, \bibinfo{person}{Amjad Almahairi}, \bibinfo{person}{Yasmine Babaei}, \bibinfo{person}{Nikolay Bashlykov}, \bibinfo{person}{Soumya Batra}, \bibinfo{person}{Prajjwal Bhargava}, \bibinfo{person}{Shruti Bhosale}, {et~al\mbox{.}}} \bibinfo{year}{2023}\natexlab{}.
\newblock \showarticletitle{Llama 2: Open foundation and fine-tuned chat models}.
\newblock \bibinfo{journal}{\emph{arXiv preprint arXiv:2307.09288}} (\bibinfo{year}{2023}).
\newblock


\bibitem[Treutlein et~al\mbox{.}(2024)]%
        {treutlein2024connecting}
\bibfield{author}{\bibinfo{person}{Johannes Treutlein}, \bibinfo{person}{Dami Choi}, \bibinfo{person}{Jan Betley}, \bibinfo{person}{Samuel Marks}, \bibinfo{person}{Cem Anil}, \bibinfo{person}{Roger~B Grosse}, {and} \bibinfo{person}{Owain Evans}.} \bibinfo{year}{2024}\natexlab{}.
\newblock \showarticletitle{Connecting the dots: Llms can infer and verbalize latent structure from disparate training data}.
\newblock \bibinfo{journal}{\emph{Advances in Neural Information Processing Systems}}  \bibinfo{volume}{37} (\bibinfo{year}{2024}), \bibinfo{pages}{140667--140730}.
\newblock


\bibitem[Vaswani et~al\mbox{.}(2017)]%
        {vaswani2017attention}
\bibfield{author}{\bibinfo{person}{Ashish Vaswani}, \bibinfo{person}{Noam Shazeer}, \bibinfo{person}{Niki Parmar}, \bibinfo{person}{Jakob Uszkoreit}, \bibinfo{person}{Llion Jones}, \bibinfo{person}{Aidan~N Gomez}, \bibinfo{person}{{\L}ukasz Kaiser}, {and} \bibinfo{person}{Illia Polosukhin}.} \bibinfo{year}{2017}\natexlab{}.
\newblock \showarticletitle{Attention is all you need}.
\newblock \bibinfo{journal}{\emph{Advances in neural information processing systems}}  \bibinfo{volume}{30} (\bibinfo{year}{2017}).
\newblock


\bibitem[Vijayakumar et~al\mbox{.}(2016)]%
        {vijayakumar2016diverse}
\bibfield{author}{\bibinfo{person}{Ashwin~K Vijayakumar}, \bibinfo{person}{Michael Cogswell}, \bibinfo{person}{Ramprasath~R Selvaraju}, \bibinfo{person}{Qing Sun}, \bibinfo{person}{Stefan Lee}, \bibinfo{person}{David Crandall}, {and} \bibinfo{person}{Dhruv Batra}.} \bibinfo{year}{2016}\natexlab{}.
\newblock \showarticletitle{Diverse beam search: Decoding diverse solutions from neural sequence models}.
\newblock \bibinfo{journal}{\emph{arXiv preprint arXiv:1610.02424}} (\bibinfo{year}{2016}).
\newblock


\bibitem[Wang et~al\mbox{.}(2022)]%
        {wang2022self}
\bibfield{author}{\bibinfo{person}{Xuezhi Wang}, \bibinfo{person}{Jason Wei}, \bibinfo{person}{Dale Schuurmans}, \bibinfo{person}{Quoc Le}, \bibinfo{person}{Ed Chi}, \bibinfo{person}{Sharan Narang}, \bibinfo{person}{Aakanksha Chowdhery}, {and} \bibinfo{person}{Denny Zhou}.} \bibinfo{year}{2022}\natexlab{}.
\newblock \showarticletitle{Self-consistency improves chain of thought reasoning in language models}.
\newblock \bibinfo{journal}{\emph{arXiv preprint arXiv:2203.11171}} (\bibinfo{year}{2022}).
\newblock


\bibitem[Wang et~al\mbox{.}(2024)]%
        {wang2024llm}
\bibfield{author}{\bibinfo{person}{Xinyuan Wang}, \bibinfo{person}{Liang Wu}, \bibinfo{person}{Liangjie Hong}, \bibinfo{person}{Hao Liu}, {and} \bibinfo{person}{Yanjie Fu}.} \bibinfo{year}{2024}\natexlab{}.
\newblock \showarticletitle{Llm-enhanced user-item interactions: Leveraging edge information for optimized recommendations}.
\newblock \bibinfo{journal}{\emph{arXiv preprint arXiv:2402.09617}} (\bibinfo{year}{2024}).
\newblock


\bibitem[Wei et~al\mbox{.}(2022)]%
        {wei2022chain}
\bibfield{author}{\bibinfo{person}{Jason Wei}, \bibinfo{person}{Xuezhi Wang}, \bibinfo{person}{Dale Schuurmans}, \bibinfo{person}{Maarten Bosma}, \bibinfo{person}{Fei Xia}, \bibinfo{person}{Ed Chi}, \bibinfo{person}{Quoc~V Le}, \bibinfo{person}{Denny Zhou}, {et~al\mbox{.}}} \bibinfo{year}{2022}\natexlab{}.
\newblock \showarticletitle{Chain-of-thought prompting elicits reasoning in large language models}.
\newblock \bibinfo{journal}{\emph{Advances in neural information processing systems}}  \bibinfo{volume}{35} (\bibinfo{year}{2022}), \bibinfo{pages}{24824--24837}.
\newblock


\bibitem[Wei et~al\mbox{.}(2024)]%
        {wei2024llmrec}
\bibfield{author}{\bibinfo{person}{Wei Wei}, \bibinfo{person}{Xubin Ren}, \bibinfo{person}{Jiabin Tang}, \bibinfo{person}{Qinyong Wang}, \bibinfo{person}{Lixin Su}, \bibinfo{person}{Suqi Cheng}, \bibinfo{person}{Junfeng Wang}, \bibinfo{person}{Dawei Yin}, {and} \bibinfo{person}{Chao Huang}.} \bibinfo{year}{2024}\natexlab{}.
\newblock \showarticletitle{Llmrec: Large language models with graph augmentation for recommendation}. In \bibinfo{booktitle}{\emph{Proceedings of the 17th ACM International Conference on Web Search and Data Mining}}. \bibinfo{pages}{806--815}.
\newblock


\bibitem[Wu et~al\mbox{.}(2023)]%
        {wu2023bloomberggpt}
\bibfield{author}{\bibinfo{person}{Shijie Wu}, \bibinfo{person}{Ozan Irsoy}, \bibinfo{person}{Steven Lu}, \bibinfo{person}{Vadim Dabravolski}, \bibinfo{person}{Mark Dredze}, \bibinfo{person}{Sebastian Gehrmann}, \bibinfo{person}{Prabhanjan Kambadur}, \bibinfo{person}{David Rosenberg}, {and} \bibinfo{person}{Gideon Mann}.} \bibinfo{year}{2023}\natexlab{}.
\newblock \showarticletitle{Bloomberggpt: A large language model for finance}.
\newblock \bibinfo{journal}{\emph{arXiv preprint arXiv:2303.17564}} (\bibinfo{year}{2023}).
\newblock


\bibitem[Yao et~al\mbox{.}(2023)]%
        {yao2023value}
\bibfield{author}{\bibinfo{person}{Jing Yao}, \bibinfo{person}{Xiaoyuan Yi}, \bibinfo{person}{Xiting Wang}, \bibinfo{person}{Yifan Gong}, {and} \bibinfo{person}{Xing Xie}.} \bibinfo{year}{2023}\natexlab{}.
\newblock \showarticletitle{Value fulcra: Mapping large language models to the multidimensional spectrum of basic human values}.
\newblock \bibinfo{journal}{\emph{arXiv preprint arXiv:2311.10766}} (\bibinfo{year}{2023}).
\newblock


\bibitem[Yu et~al\mbox{.}(2023a)]%
        {yu2023xsimgcl}
\bibfield{author}{\bibinfo{person}{Junliang Yu}, \bibinfo{person}{Xin Xia}, \bibinfo{person}{Tong Chen}, \bibinfo{person}{Lizhen Cui}, \bibinfo{person}{Nguyen Quoc~Viet Hung}, {and} \bibinfo{person}{Hongzhi Yin}.} \bibinfo{year}{2023}\natexlab{a}.
\newblock \showarticletitle{XSimGCL: Towards extremely simple graph contrastive learning for recommendation}.
\newblock \bibinfo{journal}{\emph{IEEE Transactions on Knowledge and Data Engineering}} \bibinfo{volume}{36}, \bibinfo{number}{2} (\bibinfo{year}{2023}), \bibinfo{pages}{913--926}.
\newblock


\bibitem[Yu et~al\mbox{.}(2022)]%
        {yu2022graph}
\bibfield{author}{\bibinfo{person}{Junliang Yu}, \bibinfo{person}{Hongzhi Yin}, \bibinfo{person}{Xin Xia}, \bibinfo{person}{Tong Chen}, \bibinfo{person}{Lizhen Cui}, {and} \bibinfo{person}{Quoc Viet~Hung Nguyen}.} \bibinfo{year}{2022}\natexlab{}.
\newblock \showarticletitle{Are graph augmentations necessary? simple graph contrastive learning for recommendation}. In \bibinfo{booktitle}{\emph{Proceedings of the 45th international ACM SIGIR conference on research and development in information retrieval}}. \bibinfo{pages}{1294--1303}.
\newblock


\bibitem[Yu et~al\mbox{.}(2023b)]%
        {yu2023self}
\bibfield{author}{\bibinfo{person}{Junliang Yu}, \bibinfo{person}{Hongzhi Yin}, \bibinfo{person}{Xin Xia}, \bibinfo{person}{Tong Chen}, \bibinfo{person}{Jundong Li}, {and} \bibinfo{person}{Zi Huang}.} \bibinfo{year}{2023}\natexlab{b}.
\newblock \showarticletitle{Self-supervised learning for recommender systems: A survey}.
\newblock \bibinfo{journal}{\emph{IEEE Transactions on Knowledge and Data Engineering}} \bibinfo{volume}{36}, \bibinfo{number}{1} (\bibinfo{year}{2023}), \bibinfo{pages}{335--355}.
\newblock


\bibitem[Yuan et~al\mbox{.}(2024a)]%
        {yuan2024ptf}
\bibfield{author}{\bibinfo{person}{Wei Yuan}, \bibinfo{person}{Chaoqun Yang}, \bibinfo{person}{Liang Qu}, \bibinfo{person}{Nguyen Quoc~Viet Hung}, \bibinfo{person}{Guanhua Ye}, {and} \bibinfo{person}{Hongzhi Yin}.} \bibinfo{year}{2024}\natexlab{a}.
\newblock \showarticletitle{PTF-FSR: A Parameter Transmission-Free Federated Sequential Recommender System}.
\newblock \bibinfo{journal}{\emph{ACM Transactions on Information Systems}} (\bibinfo{year}{2024}).
\newblock


\bibitem[Yuan et~al\mbox{.}(2024b)]%
        {yuan2024fellas}
\bibfield{author}{\bibinfo{person}{Wei Yuan}, \bibinfo{person}{Chaoqun Yang}, \bibinfo{person}{Guanhua Ye}, \bibinfo{person}{Tong Chen}, \bibinfo{person}{Nguyen Quoc~Viet Hung}, {and} \bibinfo{person}{Hongzhi Yin}.} \bibinfo{year}{2024}\natexlab{b}.
\newblock \showarticletitle{FELLAS: Enhancing Federated Sequential Recommendation with LLM as External Services}.
\newblock \bibinfo{journal}{\emph{ACM Transactions on Information Systems}} (\bibinfo{year}{2024}).
\newblock


\bibitem[Yuan et~al\mbox{.}(2023)]%
        {yuan2023go}
\bibfield{author}{\bibinfo{person}{Zheng Yuan}, \bibinfo{person}{Fajie Yuan}, \bibinfo{person}{Yu Song}, \bibinfo{person}{Youhua Li}, \bibinfo{person}{Junchen Fu}, \bibinfo{person}{Fei Yang}, \bibinfo{person}{Yunzhu Pan}, {and} \bibinfo{person}{Yongxin Ni}.} \bibinfo{year}{2023}\natexlab{}.
\newblock \showarticletitle{Where to go next for recommender systems? id-vs. modality-based recommender models revisited}. In \bibinfo{booktitle}{\emph{Proceedings of the 46th International ACM SIGIR Conference on Research and Development in Information Retrieval}}. \bibinfo{pages}{2639--2649}.
\newblock


\bibitem[Zhang et~al\mbox{.}(2024a)]%
        {zhang2024generative}
\bibfield{author}{\bibinfo{person}{An Zhang}, \bibinfo{person}{Yuxin Chen}, \bibinfo{person}{Leheng Sheng}, \bibinfo{person}{Xiang Wang}, {and} \bibinfo{person}{Tat-Seng Chua}.} \bibinfo{year}{2024}\natexlab{a}.
\newblock \showarticletitle{On generative agents in recommendation}. In \bibinfo{booktitle}{\emph{Proceedings of the 47th international ACM SIGIR conference on research and development in Information Retrieval}}. \bibinfo{pages}{1807--1817}.
\newblock


\bibitem[Zhang et~al\mbox{.}(2023)]%
        {zhang2023instruction}
\bibfield{author}{\bibinfo{person}{Shengyu Zhang}, \bibinfo{person}{Linfeng Dong}, \bibinfo{person}{Xiaoya Li}, \bibinfo{person}{Sen Zhang}, \bibinfo{person}{Xiaofei Sun}, \bibinfo{person}{Shuhe Wang}, \bibinfo{person}{Jiwei Li}, \bibinfo{person}{Runyi Hu}, \bibinfo{person}{Tianwei Zhang}, \bibinfo{person}{Fei Wu}, {et~al\mbox{.}}} \bibinfo{year}{2023}\natexlab{}.
\newblock \showarticletitle{Instruction tuning for large language models: A survey}.
\newblock \bibinfo{journal}{\emph{arXiv preprint arXiv:2308.10792}} (\bibinfo{year}{2023}).
\newblock


\bibitem[Zhang et~al\mbox{.}(2024b)]%
        {zhang2024benchmarking}
\bibfield{author}{\bibinfo{person}{Tianyi Zhang}, \bibinfo{person}{Faisal Ladhak}, \bibinfo{person}{Esin Durmus}, \bibinfo{person}{Percy Liang}, \bibinfo{person}{Kathleen McKeown}, {and} \bibinfo{person}{Tatsunori~B Hashimoto}.} \bibinfo{year}{2024}\natexlab{b}.
\newblock \showarticletitle{Benchmarking large language models for news summarization}.
\newblock \bibinfo{journal}{\emph{Transactions of the Association for Computational Linguistics}}  \bibinfo{volume}{12} (\bibinfo{year}{2024}), \bibinfo{pages}{39--57}.
\newblock


\bibitem[Zhao et~al\mbox{.}(2022)]%
        {zhao2022revisiting}
\bibfield{author}{\bibinfo{person}{Wayne~Xin Zhao}, \bibinfo{person}{Zihan Lin}, \bibinfo{person}{Zhichao Feng}, \bibinfo{person}{Pengfei Wang}, {and} \bibinfo{person}{Ji-Rong Wen}.} \bibinfo{year}{2022}\natexlab{}.
\newblock \showarticletitle{A revisiting study of appropriate offline evaluation for top-N recommendation algorithms}.
\newblock \bibinfo{journal}{\emph{ACM Transactions on Information Systems}} \bibinfo{volume}{41}, \bibinfo{number}{2} (\bibinfo{year}{2022}), \bibinfo{pages}{1--41}.
\newblock


\bibitem[Zheng et~al\mbox{.}(2023)]%
        {zheng2023judging}
\bibfield{author}{\bibinfo{person}{Lianmin Zheng}, \bibinfo{person}{Wei-Lin Chiang}, \bibinfo{person}{Ying Sheng}, \bibinfo{person}{Siyuan Zhuang}, \bibinfo{person}{Zhanghao Wu}, \bibinfo{person}{Yonghao Zhuang}, \bibinfo{person}{Zi Lin}, \bibinfo{person}{Zhuohan Li}, \bibinfo{person}{Dacheng Li}, \bibinfo{person}{Eric Xing}, {et~al\mbox{.}}} \bibinfo{year}{2023}\natexlab{}.
\newblock \showarticletitle{Judging llm-as-a-judge with mt-bench and chatbot arena}.
\newblock \bibinfo{journal}{\emph{Advances in Neural Information Processing Systems}}  \bibinfo{volume}{36} (\bibinfo{year}{2023}), \bibinfo{pages}{46595--46623}.
\newblock


\bibitem[Zhou et~al\mbox{.}(2023)]%
        {zhou2023lima}
\bibfield{author}{\bibinfo{person}{Chunting Zhou}, \bibinfo{person}{Pengfei Liu}, \bibinfo{person}{Puxin Xu}, \bibinfo{person}{Srinivasan Iyer}, \bibinfo{person}{Jiao Sun}, \bibinfo{person}{Yuning Mao}, \bibinfo{person}{Xuezhe Ma}, \bibinfo{person}{Avia Efrat}, \bibinfo{person}{Ping Yu}, \bibinfo{person}{Lili Yu}, {et~al\mbox{.}}} \bibinfo{year}{2023}\natexlab{}.
\newblock \showarticletitle{Lima: Less is more for alignment}.
\newblock \bibinfo{journal}{\emph{Advances in Neural Information Processing Systems}}  \bibinfo{volume}{36} (\bibinfo{year}{2023}), \bibinfo{pages}{55006--55021}.
\newblock


\bibitem[Zhu et~al\mbox{.}(2023)]%
        {zhu2023large}
\bibfield{author}{\bibinfo{person}{Yutao Zhu}, \bibinfo{person}{Huaying Yuan}, \bibinfo{person}{Shuting Wang}, \bibinfo{person}{Jiongnan Liu}, \bibinfo{person}{Wenhan Liu}, \bibinfo{person}{Chenlong Deng}, \bibinfo{person}{Haonan Chen}, \bibinfo{person}{Zheng Liu}, \bibinfo{person}{Zhicheng Dou}, {and} \bibinfo{person}{Ji-Rong Wen}.} \bibinfo{year}{2023}\natexlab{}.
\newblock \showarticletitle{Large language models for information retrieval: A survey}.
\newblock \bibinfo{journal}{\emph{arXiv preprint arXiv:2308.07107}} (\bibinfo{year}{2023}).
\newblock


\end{thebibliography}

\end{document}